\documentclass[11pt,a4paper]{article}

\usepackage{amsthm}
\usepackage{amsmath,amssymb}
\usepackage{epsfig,graphicx}
\usepackage{subfigure}
\usepackage{graphicx}
\usepackage{rotating}
\usepackage{cancel}
\usepackage{bm}
\usepackage{color}
\usepackage{comment}
\usepackage{cite}
\usepackage{psfrag}
\usepackage{slashed}
\usepackage{soul}
\usepackage{tikz}
\usepackage{hyperref}
\usepackage{bbold}
\usepackage[normalem]{ulem}

\renewcommand{\Re}{\mathrm{Re}}

\renewcommand\[{\left[}

\newcommand{\exclude}[1]{}

\def\beq{\begin{equation}}
\def\eeq{\end{equation}}

\begin{document}
\title{\LARGE{\textbf{Enhanced Axion-wind near Earth's Surface}}}
\author{ Yeray Garcia del Castillo$^{1,2}$\footnote{\href{y.garcia_del_castillo@unsw.edu.au}{y.garcia\_del\_castillo@unsw.edu.au}}, Benjamin Hammett$^1$\footnote{\href{benjamin.hammett@stud.uni-heidelberg.de}{benjamin.hammett@stud.uni-heidelberg.de}}, and Joerg Jaeckel$^{1}\footnote{\href{jjaeckel@thphys.uni-heidelberg.de}{jjaeckel@thphys.uni-heidelberg.de}}$\\[2ex]
\small{\em $^1$Institut f\"ur theoretische Physik, Universit\"at Heidelberg,} \\
\small{\em Philosophenweg 16, 69120 Heidelberg, Germany}\\
[0.5ex]
\small{\em $^2$School of Physics, The University of New South Wales,} \\
\small{\em Sydney NSW 2052, Australia}
\\[0.5ex]
}

\date{}
\maketitle

\begin{abstract}
\noindent
Several detection strategies for wave-like dark matter make use of gradients in the dark matter field, e.g. searches for spin-dependent derivative interactions in CASPEr-wind or experiments looking for oscillating forces. These gradients are usually suppressed by the local dark matter velocity $\sim 10^{-3}$. In this note we investigate how these gradients are modified in the presence of additional quadratic interactions of the dark matter field with ordinary matter. In this case the dark matter density and field are modified in the vicinity of Earth, affecting the detection sensitivity due to the change in the local field value at the Earth's surface but also due to the gradient of the field profile itself. We also use this opportunity to present results on the expected field profiles in presence of a non-vanishing relative velocity of the dark matter with respect to Earth. We also comment how this ameliorates the divergences that appear for certain attractive coupling values.

\end{abstract}

\newpage

\tableofcontents
\newpage


\section{Introduction}\label{sec:intro}
Over recent years the landscape of direct detection experiments for dark matter has significantly expanded by the development of a wide range of experiments looking for so-called wave-like dark matter (see~\cite{Sikivie:2006ni,Jaeckel:2010ni,Marsh:2015xka,Hui:2021tkt} for reviews on some wave-like dark matter candidates and~\cite{Antypas:2022asj,Adams:2022pbo} giving an overview of the experimental landscape). 

Like all direct detection experiments, also those for wave-like dark matter aim at the detection of dark matter on Earth or at least in its immediate vicinity. Therefore the properties of dark matter, e.g. its density in this environment are crucial for their sensitivity. Previous work~\cite{Hees:2018fpg,Banerjee:2022sqg,Bauer:2024yow} has already considered the effect of a quadratic coupling of the dark matter boson to matter on the dark matter density at the Earth's surface. Our aim is to also consider field gradients that are measured in certain dark matter experiments such as CASPEr-Wind~\cite{JacksonKimball:2017elr}, QUAX ~\cite{QUAX:2020adt} or accelerometers~\cite{Adelberger:2009zz,Graham:2015ifn}. 
In doing so we also account for the effects of a non-vanishing dark matter velocity, both on the gradient but also on the density.

Wave-like dark matter candidates are bosonsic and characterized by their low masses, leading to very high occupation numbers and long, i.e. macroscopic coherence length. This allows for a description in terms of classical fields~\cite{Sikivie:1983ip,Arias:2012az,Antypas:2022asj,Adams:2022pbo}. Therefore, we will solve the equations of motion for a scalar field coupled quadratically to ordinary matter in the presence of an object like Earth. 

Static, non-trivial solutions have received significant attention~\cite{Hook:2017psm,Huang:2018pbu,Zhang:2021mks,Balkin:2022qer,Balkin:2023xtr,Madge:2024aot,Gomez-Banon:2024oux}, yielding, among others, changes in the measurable values of dipole moments in the vicinity of Earth~\cite{Hook:2017psm}, effects on the neutron star equation of state~\cite{Balkin:2023xtr,Gomez-Banon:2024oux,Kumamoto:2024wjd} or the orbits of binary neutron stars~\cite{Hook:2017psm,Huang:2018pbu,Zhang:2021mks}. However, to study the effects on dark matter we are interested in solutions that are oscillating in time and also non-vanishing at large distances from the object. Stationary, oscillating, spherically symmetric solutions around an object have already been considered in~\cite{Hees:2018fpg,Banerjee:2022sqg,Bauer:2024yow} and it has been shown that stationary solutions are quickly approached in relevant experimental situations~\cite{Burrage:2024mxn}. Therefore, we consider stationary solutions, but extend them to the case of a non-vanishing dark matter velocity with respect to Earth\footnote{This breaks the spherical symmetry.}.
This is necessary for two reasons. First, non-vanishing velocities and therefore momenta are the source of the gradients that are usually considered for experiments sensitive to gradient interactions. We therefore want to compare the gradients due to the velocity to those due to the non-trivial field profile that is caused by the Earth in combination with the quadratic coupling.
Second, if the kinetic energy of the incoming particles is comparable to the potential caused by the quadratic interactions with Earth, the density and in consequence also the gradients can be significantly changed compared to the case which neglects the velocity. In this context we note that this latter effect also affects experiments that are sensitive directly to the field value (and not its gradient). Indeed, as a side effect of the non-vanishing velocity solutions, we also gain insight into the divergent solutions that have been observed  for attractive interaction in~\cite{Hees:2018fpg,Bauer:2024yow} for specific values of the coupling, reducing the problem. 

Concretely, we consider a scenario where the dark matter particle is described by a light scalar field, $\phi$, which is quadratically coupled to matter. The action for this system is given by
\begin{equation}
    S = \int d^4x \left(  \frac{1}{2} (\partial \phi)^2 -\frac{1}{2} m^2 \phi^2 -  \frac{\lambda}{2} \phi^2 \rho \right)~.
    \label{action}
\end{equation}
The quadratic coupling $\lambda$ has mass dimension $-2$ and, for simplicity, we consider a universal coupling to the energy density, $\rho$, of ordinary matter (such as Earth). This neglects equivalence principle violating effects (cf.,~\cite{Hees:2018fpg}) but those can be easily included by allowing couplings to the various species. 
At an approximate level a coupling to the matter density is a good approximation for a nucleon coupling (protons, neutrons). Specifically, we have in mind a coupling as one would expect for gluon-coupled ALPs (axion-like particles)~\cite{Hook:2017psm},
\begin{equation}
    \lambda_{\rm ALP} = - \epsilon \frac{0.059\,{\rm GeV}}{m_N}\frac{1}{4f_a^2}~,
\end{equation}
where $m_N$ is the nucleon mass and $f_{a}$ is the ``axion-decay constant'' of the gluon ($G^{\mu\nu}$) coupled ALP, ${\mathcal{L}}\supset \frac{\phi}{f_{a}} \frac{\alpha_{s}}{8\pi}G^{\mu\nu}\tilde{G}_{\mu\nu}$. Here, $\alpha_{s}$ is the strong gauge coupling constant. In this concrete model $\epsilon\approx +1$ and the coupling is attractive. To be more general we, however, also consider $\epsilon=-1$, i.e. a repulsive coupling, while still characterizing the strength by $f_{a}$.

As already mentioned we can treat the problem in classical field theory and therefore solve the equation of motion,
\begin{equation}
    \left(\Box + m^2+\lambda\rho \right)\phi = 0~.
    \label{EOM-scalar}
\end{equation}

We follow a step by step approach of including different effects. In the next section~\ref{sec:twolayer} we start with the case of vanishing relative velocity with respect to Earth. We first treat Earth as a single homogeneous sphere but then consider a two layer model (cf. also Ref.~\cite{Hees:2018fpg}) that allows us to include an atmosphere. Effects from a single, fixed velocity are included in section~\ref{sec:movingsolutions}. In section~\ref{sec:power} we then consider a whole velocity spectrum and also discuss the corresponding observable, the power spectrum. We also examine the issue of potential divergences in the case of attractive couplings. A brief summary and conclusions are presented in section~\ref{sec:conclusions}.

\section{Gradients for vanishing dark matter velocity}
\label{sec:twolayer}
\subsection{Solutions for dark matter around Earth}
To start we will consider the case where the velocity of the dark matter field vanishes. Gradients in the field are then entirely caused by the field profile that develops due to the quadratic interaction with Earth.
The vanishing velocity of the dark matter (at infinity) with respect to Earth is implemented by the following homogeneous boundary condition at infinity,
\begin{equation}
    \lim_{r \to\infty}\phi(\vec{r},t)=\phi_{0}\cos(mt+\delta)~.
\end{equation}

The homogeneity of the system is broken by the presence of Earth.
For simplicity we will model Earth as a spherical object with two layers of fixed, constant density. One representing Earth beneath the experiment and the other the atmosphere above. This is clearly still very simplistic, but we nevertheless think it already captures important aspects of the behavior.
Concretely we set,
\begin{equation}
\label{2layerDensity}
   \rho = \Bigg\{
   \begin{array}{ll}
     \rho_1 & \qquad \,\,\,\,0 \le r \le R_1 \\
         \rho_2 &\qquad R_1 \le r \le R_2 \\
          0 &\qquad  R_2 \le r
    \end{array}\, .
\end{equation}
$R_1$ is the radius up to the Earth's surface and $\rho_{1}$ its density. For the latter we take the following average value~\cite{wikiearth},
\begin{equation}
\rho_{1}=5.51 \frac{\rm{g}}{{\rm cm}^3}~.
\end{equation}
To study the dependence on the atmospheric parameters we use different average and extreme models, whose radius $R_2$ and density $\rho_2$ are given in the labels of Fig.~\ref{fig:M_rho}.

With these inputs and boundary conditions we can now solve the equation of motion Eq.~\eqref{EOM-scalar}. This has previously been done in~\cite{Hees:2018fpg} but we recall the essentials for the benefit of being self-contained.
We can separate the field into a temporal and spatial part,
\begin{equation}
    \phi(r,t) = X(r)T(t)~.
\end{equation}
The temporal equation of motion is simply a harmonic oscillator whose behavior is dictated by the boundary at infinity. The solution is given by
\begin{equation}
T(t)=\phi_{0}\cos(mt+\delta)~,    
\end{equation}
where we have already included the asymptotic field value $\phi_{0}$ for convenience.

The solutions for the spatial part are given according to their respective regions.
Inside the first layer, the solutions that stay finite at the center of Earth are given by~\cite{Hees:2018fpg}
\begin{equation}
\label{In1stLayer}
    X(r) = \Bigg\{
\begin{array}{ll}
    B\frac{\sin(r\gamma_1)}{r} & \lambda_1 < 0\\
    B\frac{\sinh(r\gamma_1)}{r} &  \lambda_1 > 0\\
\end{array}\,~,
\end{equation}
\begin{equation}
    \gamma_i = \sqrt{|\lambda_i|\rho_i}\sim\frac{\sqrt{\rho_{i}}}{f_{a}}~.
\end{equation}
Here, we introduce the shorthand $\gamma$ to encode the entire dependence on the coupling and density. 

Similarly, the solutions inside the second layer are~\cite{Hees:2018fpg}
\begin{equation}
\label{In2ndLayer}
    X(r) = \Bigg\{
    \begin{array}{ll}
    C\frac{\sin{(\gamma_2 r)}}{r} + D\frac{\cos{(\gamma_2 r)}}{r} & \lambda_2 < 0\\
    C\frac{\exp{(\gamma_2 r)}}{r} + D\frac{\exp{(-\gamma_2 r)}}{r} & \lambda_2 > 0
    \end{array}\,~.
\end{equation}
Finally, the solution outside of the atmosphere is independent of the sign of the coupling and is given by,
\begin{equation}\label{OutsideBothLayers}
    X(r) = 1-\frac{A}{r}~.
\end{equation}
The four constants A,B,C,D can be determined by the continuity conditions of the field and the gradient at the boundaries $R_1$ and $R_2$ (see also~\cite{Hees:2018fpg}). The explicit constants are given in 
Appendix~\ref{app:continuity}, notably Eqs.~\eqref{eq:twolayerrep} to \eqref{eq:dequation}. \footnote{We note a typo in the appendix of~\cite{Hees:2018fpg} where the continuity constant of the outside two layer solution is given. This is the constant A in \eqref{eq:twolayeratt}.}

It is straightforward to check that if the atmosphere layer is removed by setting $R_{1}=R_{2}$ the solution for a simple homogeneous sphere is recovered~\cite{Hees:2018fpg}.
The corresponding solutions are simply given by directly connecting \eqref{In1stLayer} and \eqref{OutsideBothLayers}. This requires two constants.
For convenience we provide them again in Appendix~\ref{app:continuity} (Eq.~\eqref{eq:onelayerrep}, Eq.~\eqref{eq:onelayeratt}).

The main physics can already be understood from  Eqs.~\eqref{In1stLayer} and \eqref{In2ndLayer}.
We can see that the repulsive interactions feature an exponential behavior whereas the attractive solutions oscillate w.r.t. their position and the coupling. 
This is in line with a simple quantum mechanical picture of Earth providing a non-trivial potential well for the dark matter particles. 
In the repulsive case Earth leads to a potential barrier. The dark matter particles that are at rest far outside Earth can only penetrate the barrier with an exponentially suppressed amplitude. In the attractive case the particles have a non-vanishing velocity inside leading to oscillations and possibly bound states. 

In both, the repulsive and attractive case, we have a non-trivial spatial dependence and therefore non-vanishing gradients. Hence, experiments that probe such gradients will become sensitive even in the absence of a dark matter velocity. Moreover, and as we will see, in some regions of parameter space these gradients are also parametrically bigger than those caused by the velocity.

\subsection{Effects on the gradients}
In absence of quadratic interactions the gradient of the dark matter field is determined by its velocity with respect to Earth. With the gradient being the momentum operator we schematically have $v\sim \nabla/m$.
To facilitate a comparison with this simple case we introduce an effective velocity on the surface of the Earth as a measure of the field gradient. To do this we take the absolute value of the field gradient, normalised by the field at infinity and the axion mass,
\begin{equation}
\label{eq:veff}
 v_{\rm eff}=\frac{|\nabla\phi|}{m\phi(\infty)}~. 
\end{equation}
This definition and normalization is chosen to facilitate an easy comparison to the simple case of a dark matter velocity without any quadratic interactions. In this case the gradient of the field scales as $\nabla\phi\sim mv\phi(\infty)$, where $v\sim 10^{-3}$ is the velocity of the dark matter.
It is noteworthy, that this effective velocity is not bounded by the speed of light, and is to be understood as a normalised gradient.

\begin{figure}[t]
    \centering
    \includegraphics[width = 0.48 \linewidth]{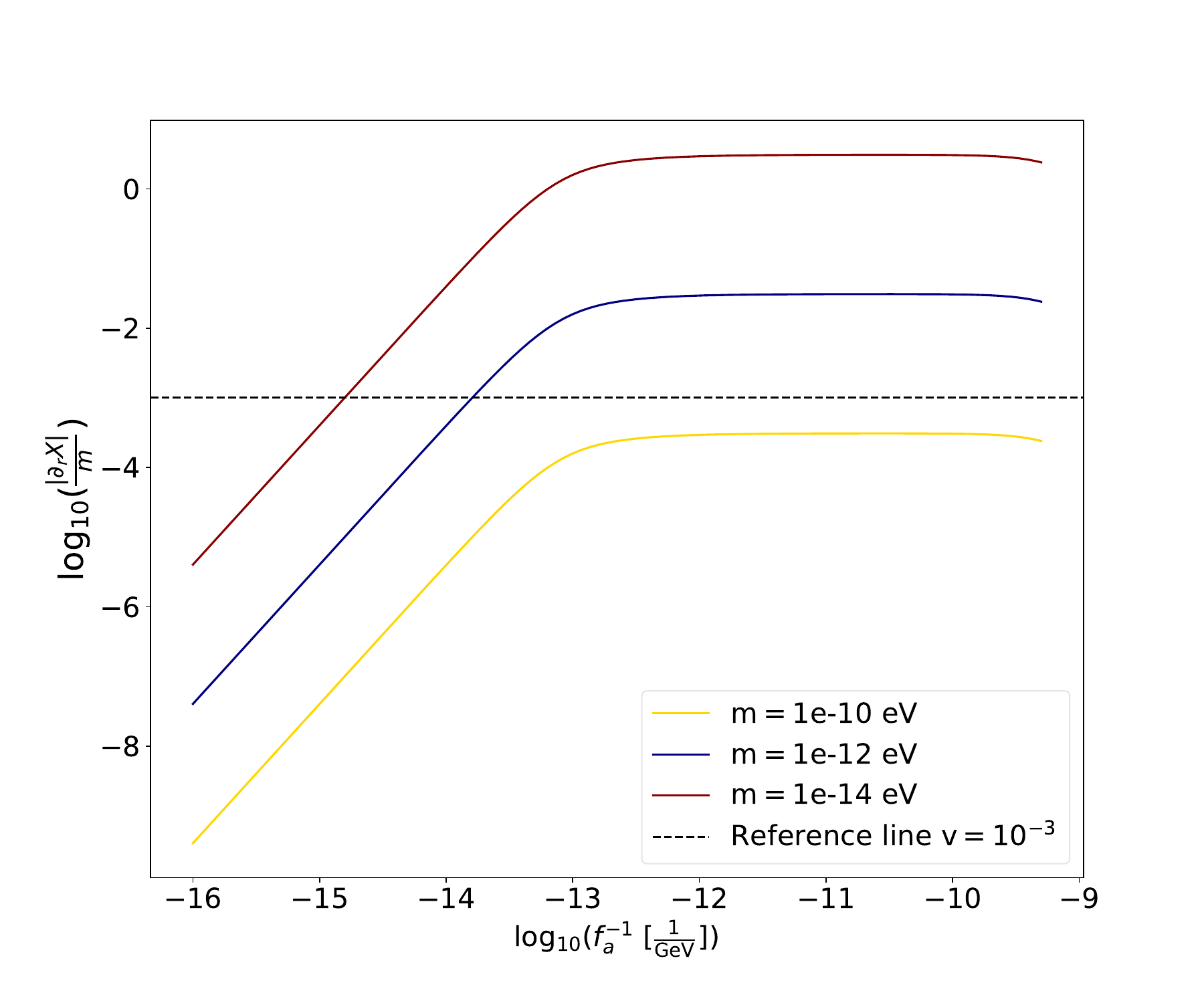}
    \includegraphics[width = 0.48 \linewidth]{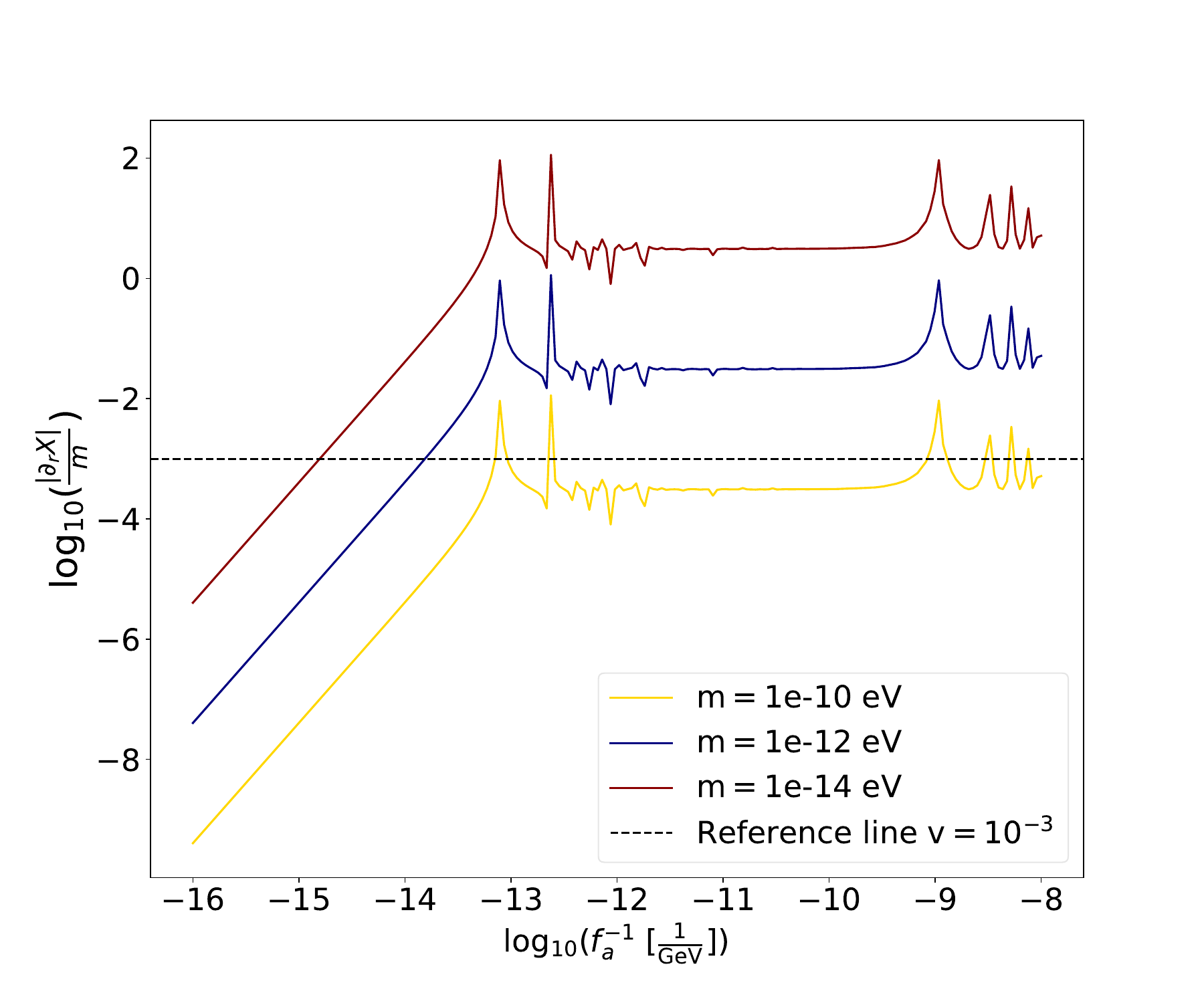}
    \caption{Effective velocity for repulsive (left panel) and attractive (right panel) coupling. The colors denote different masses as indicated in the insert. The parameters of the atmosphere were taken to be $R_{2}=\mathrm{100\, km}$ and $\rho_{2}= 0.125\,\mathrm{\frac{kg}{m^3}}$. As a reference we indicate the expected order of magnitude of the gradients in absence of quadratic interactions with the black dashed line a velocity of $10^{-3}$. For the attractive case the region to the right of the first peak should be treated with care (see main text). } 
    \label{fig:P_v(f)}
\end{figure}

The effects of the quadratic interaction can be seen in Fig.~\ref{fig:P_v(f)} where we show the effective velocity on the surface of the Earth as a function of the inverse scale $\frac{1}{f_a}$. 
On a broad qualitative level the curves for attractive and repulsive couplings show fairly similar behavior. Starting from low couplings the effective velocity increases and then plateaus towards larger values of the coupling. 

The effective velocity increases linearly with decreasing mass. This is because the spatial part of the stationary solutions actually does not depend on the mass and therefore the effective velocity of Eq.~\eqref{eq:veff} linearly increases with $\sim 1/m$. Importantly, at low masses an enhancement of many orders of magnitude compared to the situation without quadratic interactions is possible (see the black dashed line indicating the expectation without quadratic interactions). 
This is the main result of this section.

At this point we should also discuss the peaks visible in the case of attractive interactions. For attractive interactions Earth provides a potential well for the dark matter particles. If this well is sufficiently deep and large it features a number of bound states. With increasing coupling more and more bound states become possible. The peaks arise whenever an additional bound state is just becoming possible. For vanishing incoming velocities these ``resonances'' lead to infinities in the field value signalling a breakdown of the approximation and modelling assumptions used, as already noted in~\cite{Hees:2018fpg} and discussed in some additional detail for axion-like particles in~\cite{Bauer:2024yow}. We note here that for non-vanishing velocity the results are finite (see the next section~\ref{sec:movingsolutions}).
Nevertheless, when integrating over a full velocity distribution of dark matter it is likely that there is always a part the velocity spectrum of dark matter that leads to a vanishing relative velocity, which may be problematic.  We will discuss this in more detail in section~\ref{sec:power}.

\begin{figure}[!t]
    \centering
    \includegraphics[width = 0.48\linewidth]{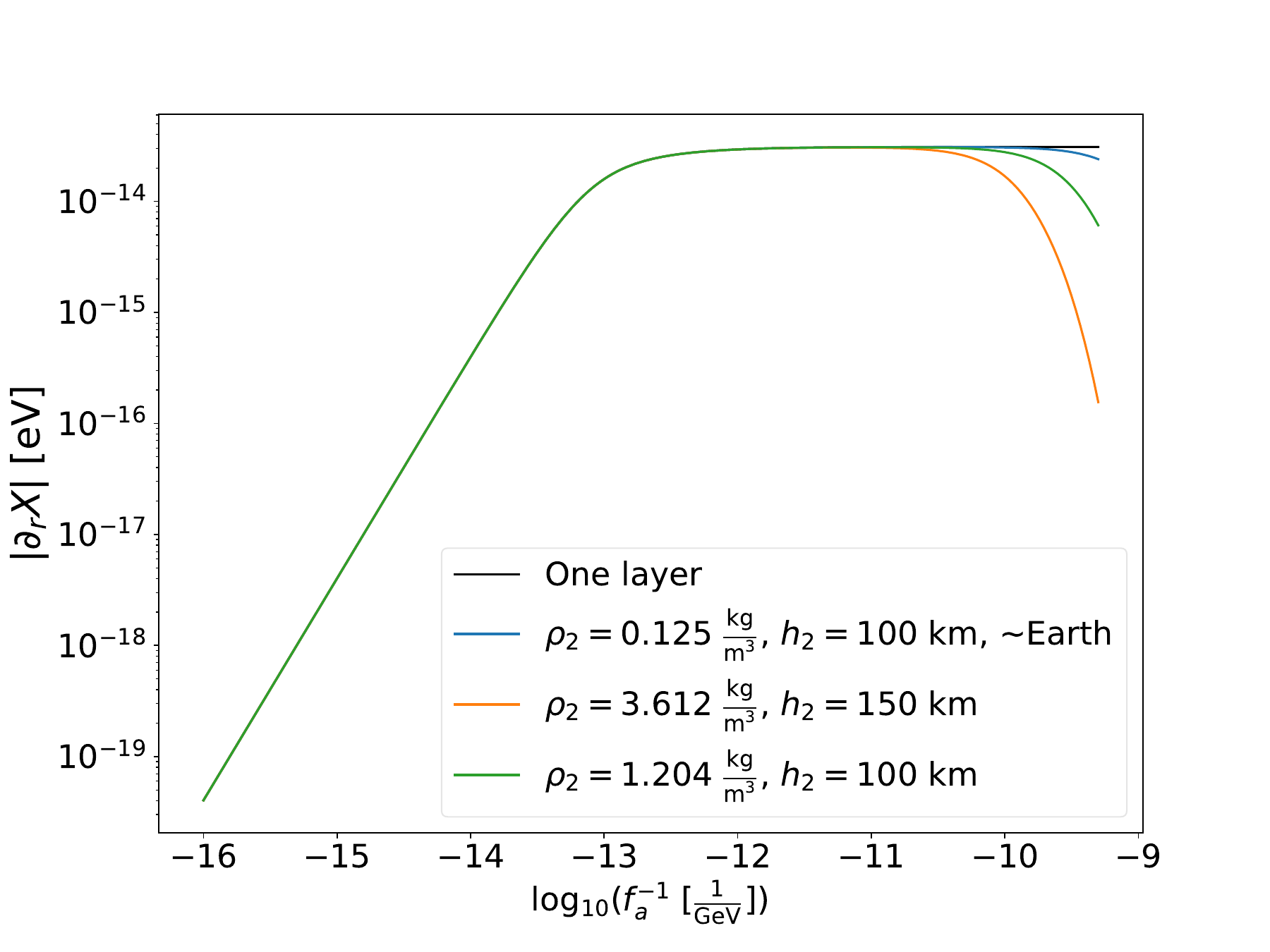}
    \includegraphics[width = 0.48\linewidth]{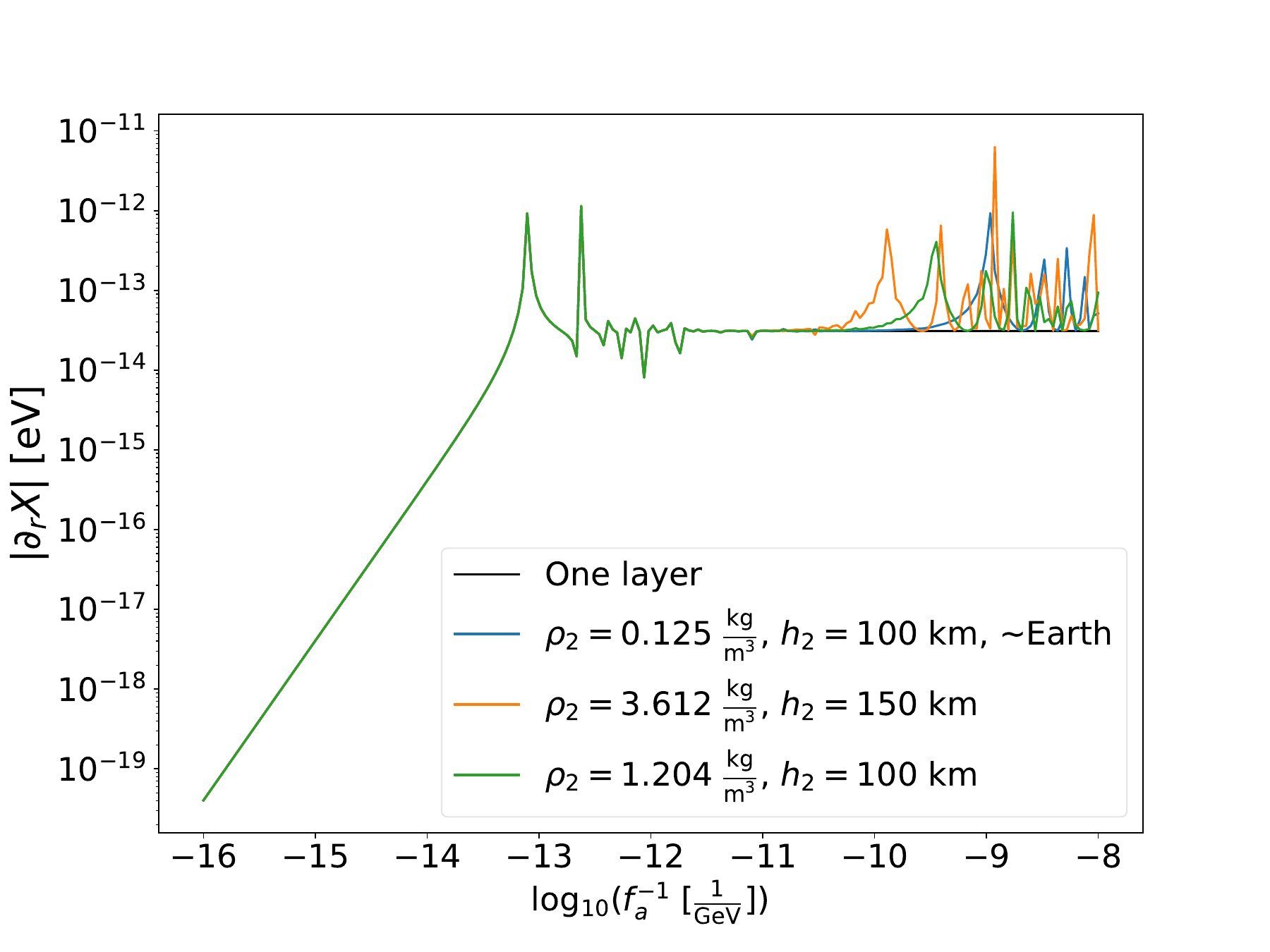}
    \caption{Value of the gradient normalized to the field at infinity for different atmospheric models. The left panel is for repulsive couplings and the right panel for attractive ones. The inserts show the atmospheric parameters used for the different lines. The parameter values marked with $\sim {\rm Earth}$ are the closest to the actual Earth atmosphere amongst the simplistic models and are the ones used in Fig.~\ref{fig:P_v(f)}.}
    \label{fig:M_rho}
\end{figure}

Let us now take a closer look at the effect of the atmosphere.
For repulsive couplings we expect that the atmosphere suppresses the field and therefore also the gradient at large couplings. In physical terms the repulsive interactions make it harder for the dark matter particles to penetrate the atmosphere and reach the surface where we perform our measurements. This can already be glimpsed from the high coupling end of Fig.~\ref{fig:P_v(f)} but seen much better in the left panel of Fig.~\ref{fig:M_rho} where we show results for different atmospheric models.
The effect of the atmosphere is usually quite small for couplings $1/f_{a}\lesssim 10^{-10}\,{\rm GeV}^{-1}$.

For attractive couplings there is no clear suppression due to the atmosphere. However, there, too, the effect usually is only important for couplings $1/f_{a}\gtrsim 10^{-10}\,{\rm GeV}^{-1}$. In any case at this point we are in a region where the approximation and modelling are questionable.

\section{Non-vanishing dark matter velocity}
\label{sec:movingsolutions}
So far, we have discussed the non-moving solutions for the dark matter field, that correspond to a homogeneous boundary condition given by $\lim_{r \to \infty} \phi(\vec{r},t)=\phi_{0}\cos(mt+\delta)$. As already mentioned this corresponds to a vanishing relative velocity between Earth and dark matter particles.
However, in the Milky Way halo we expect that dark matter particles move with a significant velocity of the order of $\sim 10^{-3}$. Moreover, Earth moves with respect to this halo towards Cygnus at a velocity of $v_{0} \sim 10^{-3}$~\cite{disp}. In absence of quadratic interactions the resulting relative velocity with respect to the experiment is responsible for the gradients. For a comparison between the coupling induced gradients and those from the original dark matter velocity it is only natural to take the initial velocity into account.

Indeed there is a second reason to include velocities. As already mentioned the presence of Earth together with the quadratic interactions induces an effective potential for the dark matter particles. Now, for the behavior of the particles in the vicinity of Earth it is therefore clearly relevant whether this effective potential barrier/well is bigger or smaller than the kinetic energy of the dark matter particles. As the dark matter velocity is fixed (see above) we can expect that this modifies the results for larger dark matter masses where the kinetic energy cannot be neglected anymore.

Following the results of the previous section and for simplicity we neglect the Earth's atmosphere and approximate the Earth simply as a homogeneous sphere of radius $R=R_1$ and density $\rho=\rho_1$.

\subsection{Solutions for non-vanishing  velocity}
\label{nonvanishingsol}
To take into account the effect of a non-vanishing DM velocity, we can add a momentum to the previous boundary condition, leading to 
\begin{equation}
    \lim_{r \to \infty} \phi(\vec{r},t)=\phi_{0}\cos( \omega t-\vec{k}\vec{r}+\delta)~.
    \label{limitrofe}
\end{equation}
Here, $\omega=\sqrt{m^2+k^2}$ and $k\approx m v$, where we are taking the particles to be non-relativistic. In this way, the non-vanishing momentum is taken into account at the boundary while the field still fulfills the Klein-Gordon equation. 

In order to solve the differential equation in Eq. \eqref{EOM-scalar}, with this new boundary condition, we will carry out a partial wave analysis.\footnote{Indeed the problem and its solutions are essentially equivalent to scattering on a spherical constant potential in quantum mechanics. For a detailed discussion of this problem see, e.g.~\cite{lecture}.} To do so we take the boundary condition of Eq.~\eqref{limitrofe} into its complex wave form, and add a term proportional to $\sim 1/r$ in order to take into account the effect of the sphere on the incoming plane wave. Setting $\delta=0$ for simplicity, the boundary condition we will use reads
\begin{equation}
       \lim_{r \to \infty} \phi(\vec{r},t)=\phi_{0}\left(e^{-i \omega t+i \vec{k}\vec{r}}+f(\theta)\frac{e^{-i \omega t+i kr}}{r}\right)~.
\end{equation}
  
Assuming that the incident wave is in the direction of the $z$-axis, Eq~\eqref{EOM-scalar} can be solved as follows. We find two different types of solutions, depending on whether $\omega^2 -m^2 - \lambda \rho> 0$ or $\omega^2 -m^2 - \lambda \rho< 0$. 

For the $\omega^2 -m^2 - \lambda \rho> 0$ case, the solution reads
        \begin{equation}
    \phi(\vec{r},t)= 
     \begin{cases}
  \Re \left( \phi_{0} e^{-i \omega t} \sum_{l=0}^{\infty} A_l j_l(k' r)P_l(\cos(\theta)) \right)& \text{if } r<R\\
     \Re \left( \phi_{0} e^{-i \omega t}  \sum_{l=0}^{\infty}(i)^l (2l+1)(j_l(kr)+ia_l h_l^1(kr))P_l(\cos(\theta)) \right)& \text{if }  r>R\\
       \end{cases}
       \label{eq11}
       \end{equation}
where $j_l(x)$ is the spherical Bessel function of first kind, $h_l^1(x)$ the spherical Hankel function of first kind, $P_l(x)$ the Legendre polynomial of order $l$ and $k'=\sqrt{\omega^2-m^2-\lambda \rho}$. $A_l$ and $a_l$ are constants that are determined using the continuity of the field and its derivative at $r=R$. Applying these conditions, the constants are given by
 \begin{eqnarray}
i a_{l}&=&\frac{k j_{l}'(k R)j_{l}(k' R)- k'j_{l}'(k' R)j_{l}(k R)}{k' h_{l}(k R)j_{l}'(k' R)-k j_{l}(k 'R)h_{l}'(k R)}~,
\label{toneria}
\\
A_{l}&=&(i)^{l}(2l+1)\frac{j_{l}(k R)+i a_{l} h_{l}(k R)}{j_{l}(k'R)}~.
\end{eqnarray}

Note that we are taking the real part of the solutions since the field we are treating is real. For the $\omega^2 -m^2 - \lambda \rho< 0$ case, the solution reads
\begin{equation}
    \phi(\vec{r},t)= 
     \begin{cases}
   \Re \left( \phi_{0} e^{-i \omega t} \sum_{l=0}^{\infty} A_l i_l(k' r)P_l(\cos(\theta)) \right) & \text{if } r<R\\
      \Re \left(\phi_{0} e^{-i \omega t}  \sum_{l=0}^{\infty}(i)^l (2l+1)(j_l(kr)+ia_l h_l^1(kr))P_l(\cos(\theta)) \right) & \text{if }  r>R\\
     \end{cases}
     \label{eq22}
\end{equation}
where $i_l(x)$ is the modified spherical Bessel function of first kind, $k'=\sqrt{-\omega^2+m^2+\lambda \rho}$ and again $ A_l$ and $a_l$ are constants that can be determined with the continuity conditions, leading to
\begin{eqnarray}
i a_{l}&=&\frac{k j_{l}'(k R)i_{l}(k' R)- k'i_{l}'(k' R)j_{l}(k R)}{k' h_{l}(k R)i_{l}'(k' R)-k i_{l}(k 'R)h_{l}'(k R)}~,
\label{tonteria2}
\\
A_{l}&=&(i)^{l}(2l+1)\frac{j_{l}(k R)+i a_{l} h_{l}(k R)}{i_{l}(k'R)}~.
\end{eqnarray}

As a check, we show in Appendix~\ref{app:limit} that, for $v\to 0$, these solutions reduce to the ones in the case of a homogeneous boundary condition (without an atmosphere). 

Moreover, we show in Appendix~\ref{Proof} that for $k\neq 0$ these solutions, most notably the $a_{l}$, remain finite for all values of the coupling, including attractive ones. The divergences that are present for certain attractive values re-appear in the limit $k\to 0$, as discussed in Appendix~\ref{app:limit2}. We discuss a possible resolution in the next section~\ref{sec:power}.
  
Note that the solutions we have found can be expressed in general as
\begin{equation}
      \phi(\vec{r},t)=  \Re \left( e^{-i \omega t+ i\delta} \Tilde{\phi}(\vec{r}) \right)= \Tilde{\phi}(\vec{r}) \cos(\omega t- \delta) ~,
      \label{decompositon}
\end{equation}
where $\Tilde{\phi}(\vec{r})$ is the modulus of the complex solution and $\delta$ its phase. In this work we are mainly interested in the amplitude of the field and its gradient, which can be obtained from the modulus of the complex solution, $\Tilde{\phi}(\vec{r})$ and $\nabla \Tilde{\phi}(\vec{r})$ respectively.

\subsection{Effective velocity on Earth's surface}

\begin{figure}[!t]    
\centering
    \includegraphics[width = 0.48 \linewidth]{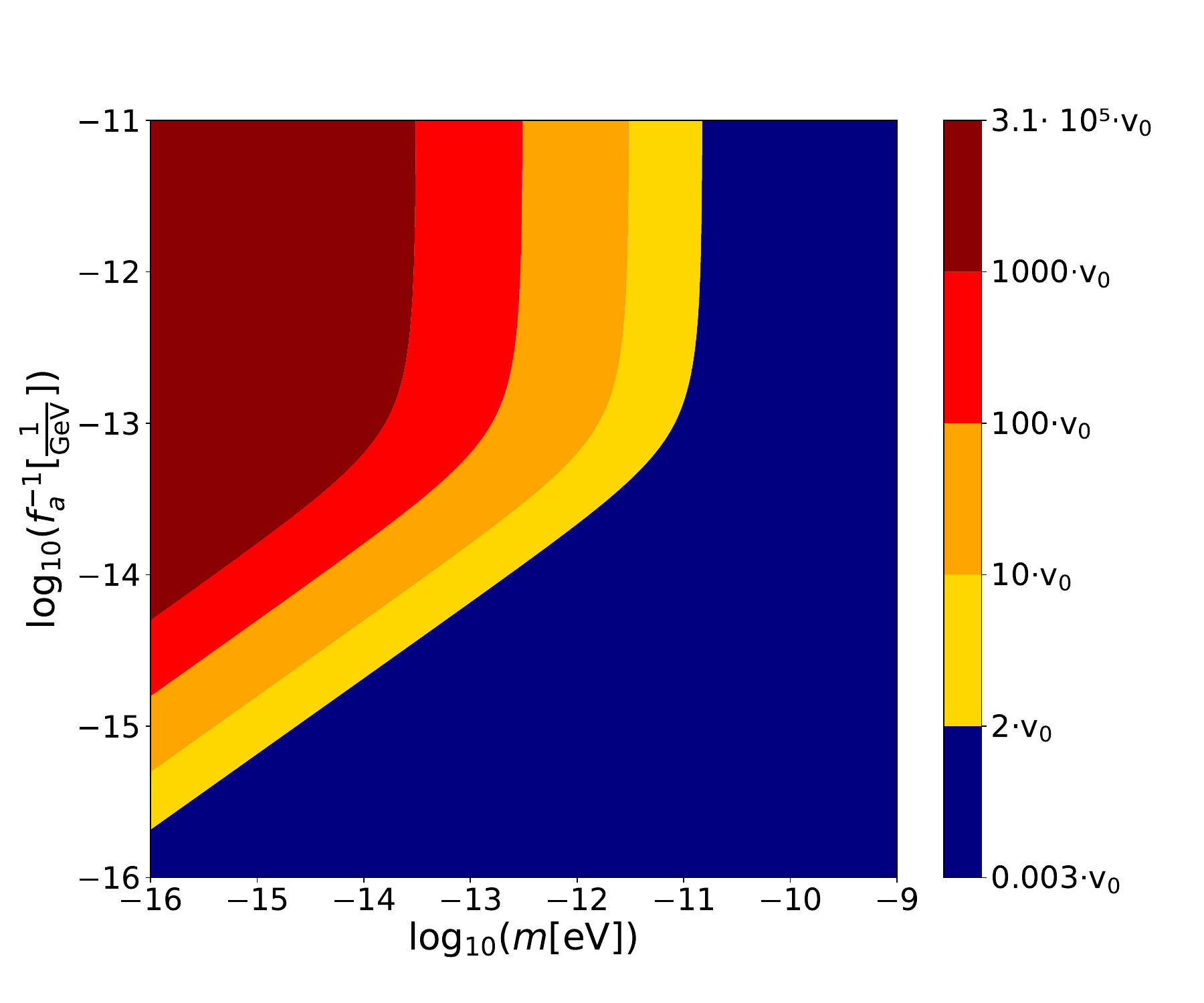}
    \includegraphics[width = 0.48 \linewidth]{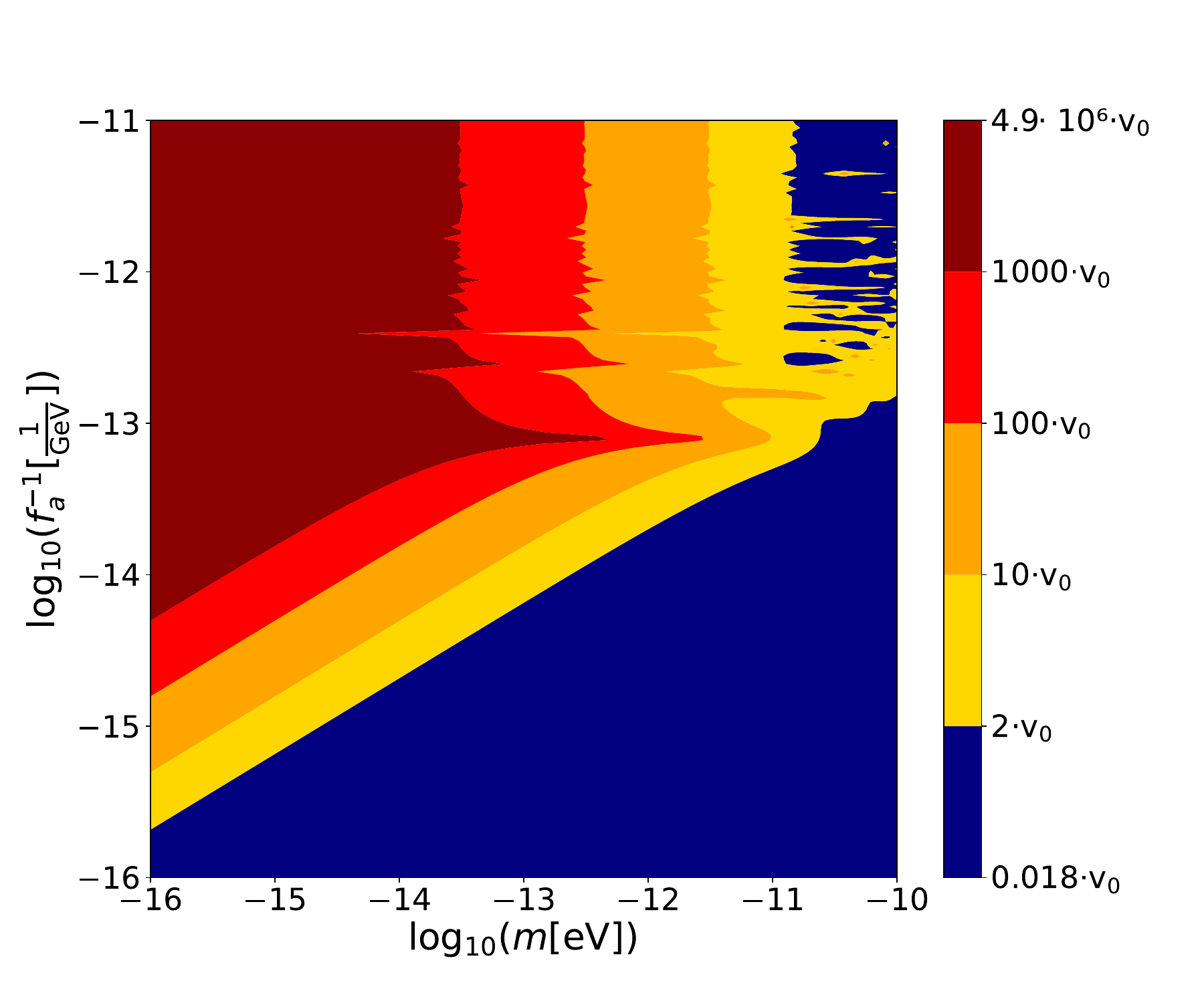}
    \caption{Regions of enhancement for the dynamic gradient. As before the left panel is for repulsive and the right panel for attractive interactions. We have chosen a single initial velocity $v_0=10^{-3}$ and a measurement point $\vec{r} = (0,0,R)$. The colourbar on the right indicates the lines at specified multiples of $v_0$. Note that we have chosen yellow for values greater than $2v_0$ to indicate a significant enhancement.} 
    \label{fig:P_enhancement}
\end{figure}

Equipped with the non-vanishing velocity solutions to the field equations we can now obtain the gradient on Earth's surface. In practice we are mostly interested in the regions where the effective velocity is higher than the initial velocity $\sim 10^{-3}$.
To do so we define an enhancement region of the parameter space, where the effective velocity surpasses $10^{-3}$, it is here that the effect of the quadratic coupling on the gradient rivals and surpasses the contribution of the relative velocity of  Earth to the dark matter.
The results are shown in Fig.~\ref{fig:P_enhancement} where the blue region indicates no or only moderate enhancement and the other colors show regions of significant enhancement (we change the color to yellow when the enhancement is at least 2).

As Fig.~\ref{fig:P_enhancement} shows, significant enhancement of the gradient is possible. As we can see the effect is most dramatic at low masses. This can be best understood by comparing the effective non-relativistic potential experienced by the dark matter particles to their kinetic energy, 
\begin{equation}
    V_{\rm{non-rel}}=\frac{\lambda \rho}{2m}~,\qquad\qquad E_{kin}=\frac{1}{2}mv^2~.
\end{equation}
The relevant ratio,
\begin{equation}
    \frac{V_{\rm non-rel}}{E_{kin}}= \frac{\lambda\rho}{m^2v^2}\sim \frac{1}{m^2}~,
\end{equation}
therefore grows rapidly with decreasing mass $m$.

\subsection{Effective field values on Earth's surface}
Having the solutions with non-vanishing velocity also allows us to improve on the estimate for the field value on the surface. This is relevant for experiments that are directly sensitive to the field value, e.g. CASPEr-Electric~\cite{JacksonKimball:2017elr} or haloscopes like ADMX \cite{admx1, admx2, admx3, admx4, admx5}, BREAD~\cite{BREAD:2021tpx}, WISPDMX \cite{ma1},  KLASH \cite{ma3} or ACTION \cite{ma4} to name a few.

\begin{figure}[!t]    
\centering
    \includegraphics[width = 0.48 \linewidth]{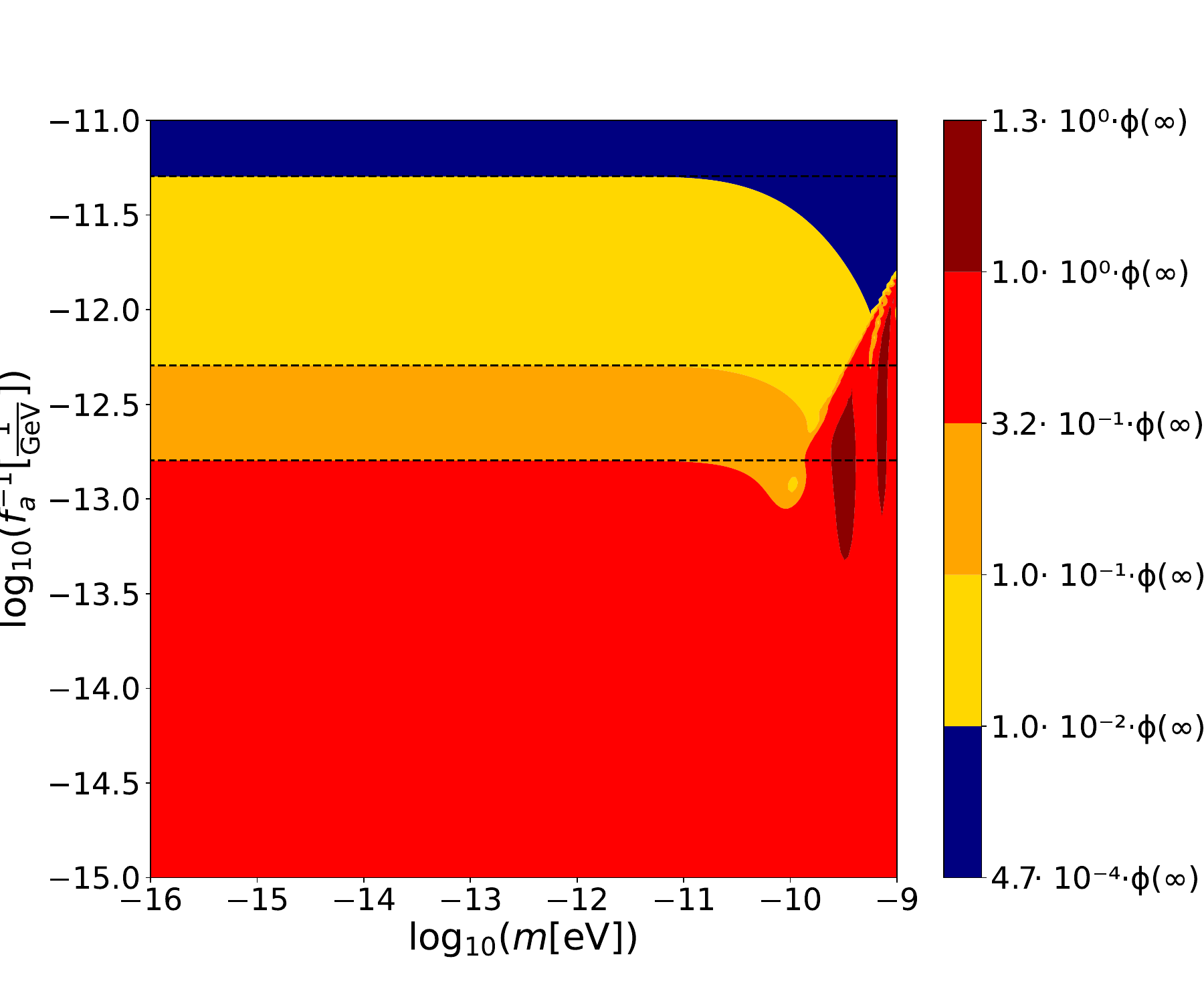}
    \includegraphics[width = 0.48 \linewidth]{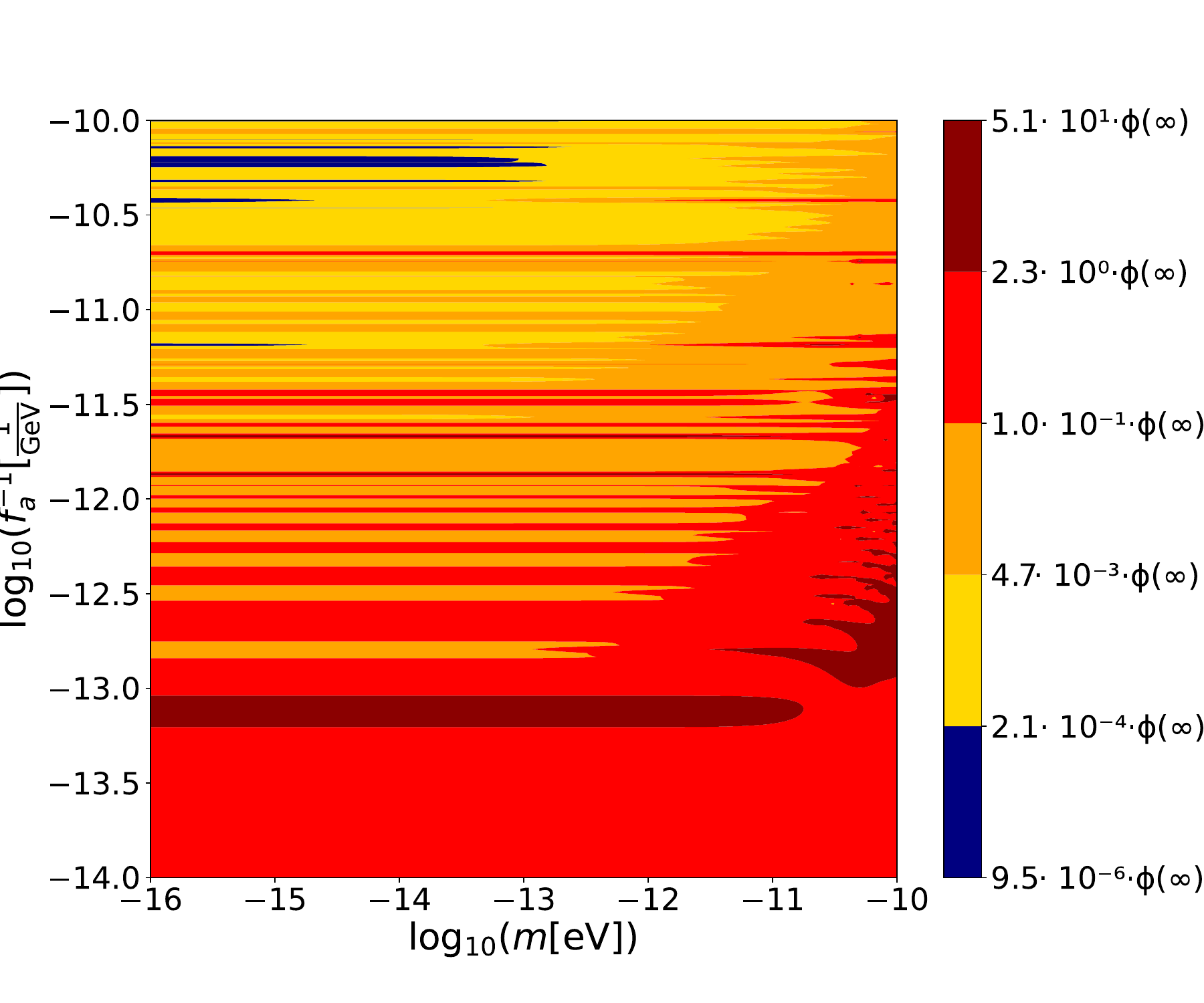}
    \caption{Modification of the absolute field value on Earth's surface compared to the value far away from Earth. As usual the left panel is for repulsive and the right panel for attractive interactions. The colors indicate the size of the ratio Eq.~\eqref{eq:phieff}. In the higher coupling regions there are many more dark red (enhanced) lines, however they are too thin for the resolution of this plot. We have chosen a single velocity $v=10^{-3}$ and a measurement point $\vec{r} = (0,0,R)$. In the repulsive plot the black lines show the ratios if we consider a vanishing incoming velocity for comparison. }
    \label{fig:fieldcomparison}
\end{figure}

As before it makes sense to compare the field value to the one in absence of quadratic interactions, i.e. the one outside Earth's sphere of influence,
\begin{equation}
\label{eq:phieff}
 r_{\rm eff}=\frac{|\phi_{\rm{Earth surface}}|}{|\phi(\infty)|}~.
\end{equation}
The resulting modifications are shown in Fig.~\ref{fig:fieldcomparison}. For low masses significant changes in the field value are possible, whereas the effect is smaller for larger masses where the kinetic energy is higher. For repulsive couplings we also compare to the case of vanishing velocities (black lines) that has already been discussed in Ref.~\cite{Banerjee:2022sqg}. Including the effects of kinetic energy the modifications of the field value are reduced for higher masses. 

Note also that for higher attractive (as well as repulsive) couplings the field value (outside the thin resonant regions) is actually often decreased. This is due to the analogous effect of quantum mechanical reflection. 

\section{Non-trivial velocity distribution and the power spectrum}
\label{sec:power}
So far, in the moving solutions, we have considered a monochromatic dark matter wave. Even though this treatment captures some essential qualitative features it is still a significant simplification. In the vicinity of Earth we expect dark matter to have a velocity distribution. In this case we have an incoming plane wave for each (vectorial) velocity. Notably each of them can have their own random phases.
If we naively average over the random phases we would get that the gradient is vanishing.  In this kind of situation it is useful to appeal to the power spectrum. 

\subsection{Calculating the power spectrum}
In the following we will compute the power spectrum using the Fourier transform of the autocorrelator (Wiener–Khinchin theorem~\cite{wiener,Khintchine1934KorrelationstheorieDS}), 
\begin{equation}
    S_{\partial_i\phi \partial_j \phi}(\omega)=\mathcal{FT}(R_{\partial_i\phi \partial_j \phi}(\tau))~,
\end{equation}
where $i$ and $j$ can be any of the coordinates.\footnote{We are mostly interested in cases $i=j$; for $i\neq j$ we have in mind a symmetrized version.} 

With the solutions from Sec. \ref{sec:movingsolutions} in mind, we can express the field solution for a monochromatic wave with momentum $\vec{k}$ and random phase $\alpha(\vec{k})$ as
\begin{equation}
     \phi_{\vec{k}}(\vec{r},t)=\Re \left(\phi_{0} e^{-i \omega(k) t+ i \alpha(\vec{k})} \sum_{l=0}^{\infty}  e^{i \delta_l(k)} \phi_l(k,r) P_l(\hat{k} \hat{r}) \right)~,
\end{equation}
where $\phi_l(k,r)$ represents the modulus of the radial part in the field solutions given by Eq. \eqref{eq11} and \eqref{eq22}, and $\delta_l(k)$ its phase. The partial derivatives are then given by
\begin{equation}
     \partial_i \phi_{\vec{k}}(\vec{r},t)=\Re \left(\phi_{0} e^{-i \omega(k) t+ i \alpha(\vec{k})} \sum_{l=0}^{\infty}  e^{i \delta_l(k)} \partial_i(\phi_l(k,r)P_l(\hat{k} \hat{r})) \right).
\end{equation}
Integrating over a general distribution in momentum the derivative of field is given by
\begin{equation}
     \partial_i \phi(\vec{r},t)=\Re \left(\phi_{0} \int \sqrt{f(\vec{k})} e^{-i \omega(k) t+ i \alpha(\vec{k})}\sum_{l=0}^{\infty}  e^{i \delta_l(k)} \partial_i(\phi_l(k,r)P_l(\hat{k} \hat{r})) \, \frac{d^3k}{(2 \pi)^3} \right).
\end{equation}
In order to simplify the notation and get rid of the real part of the expression we can define
\begin{equation}
    A(\vec{r},t)=\phi_{0} \int \sqrt{f(\vec{k})} e^{-i \omega(k) t} c_{\vec{k}} \sum_{l=0}^{\infty}  e^{i \delta_l(k)} \partial_i(\phi_l(k,r)P_l(\hat{k} \hat{r})) \,\frac{d^3k}{(2 \pi)^3}~,
\end{equation}
where $c_{\vec{k}}$ encodes now the information about the random phase. The derivatives then read
\begin{equation}
      \partial_i \phi(\vec{r},t)=\frac{A(\vec{r},t)+A(\vec{r},t)^*}{2}~.
\end{equation}
At this point we can proceed by calculating the autocorrelator that is given by
\begin{equation}
    R_{\partial_i\phi \partial_j \phi}(\tau)=<\partial_i\phi(\vec{r},t)\partial_j \phi(\vec{r},t-\tau)>~,
\end{equation}
where $<>$ refers to an ensemble average. Since the phases $\alpha(\vec{k})$ are random, when we do the averaging over the $c_{\vec{k}}$ we have the following identities (cf., e.g.~\cite{Berges:2015kfa}),
\begin{equation}
    <c_{\vec{k}} c^*_{\vec{k'}}>=(2\pi)^3\delta^3(\vec{k}-\vec{k'})
    \label{eq:average1}
\end{equation}
\begin{equation}
    <c_{\vec{k}} c_{\vec{k'}}>=<c^*_{\vec{k}} c^*_{\vec{k'}}>=0~.
    \label{eq:average2}
\end{equation}
With this in mind, we can calculate the autocorrelator,
\begin{equation}
     R_{\partial_i\phi \partial_j \phi}(\tau)= \lim_{t \to \infty}\frac{1}{2T} \int_{-T}^{T}<(\frac{A(\vec{r},t)+A(\vec{r},t)^*}{2})(\frac{A(\vec{r},t-\tau)+A(\vec{r},t-\tau)^*}{2})> dt~.
\end{equation}
Using Eq. \eqref{eq:average1} and \eqref{eq:average2}, the expression can be simplified to 
\begin{multline}
R_{\partial_i\phi \partial_j \phi}(\tau)= \frac{\phi_0^2}{2} \int \frac{d^3k}{(2\pi)^3} f(\vec{k}) \\
\sum_{ll'} \cos(\omega(k) \tau -\delta_l(k)+\delta_{l'}(k)) \partial_i(\phi_l(k,r)P_l(\hat{k} \hat{r})) \partial_j(\phi_{l'}(k,r)P_{l'}(\hat{k} \hat{r}))~,
\end{multline}
where now we have a double sum over the two indices $l$ and $l'$. Once we have computed the autocorrelator, we can calculate the power spectrum by doing the Fourier transform,
\begin{equation}
     S_{\partial_i\phi \partial_j \phi}(\omega_0)=\int_{-\infty}^{\infty}e^{-i\omega_0 \tau}R_{\partial_i\phi \partial_j \phi}(\tau)d\tau~.
\end{equation}
Using the Fourier transform of the cosine, this expression simplifies to
\begin{equation}
    \begin{aligned}
    S_{\partial_i\phi \partial_j \phi}(\omega_0) &= \frac{\phi_0^2}{4 (2\pi)^2} \int f(\vec{k})\sum_{ll'}\left(  \delta(\omega_0-\omega)e^{-i(\delta_l(k)-\delta_{l'}(k))} \right. \\
    &\qquad\qquad \left. + \delta(\omega_0+\omega)e^{i(\delta_l(k)-\delta_{l'}(k))} \right)
    \partial_i(\phi_l(k,r)P_l(\hat{k} \hat{r})) \partial_j(\phi_{l'}(k,r)P_{l'}(\hat{k} \hat{r})) d^3k~.
    \end{aligned}
    \label{powdelta}
\end{equation}
Using the symmetry in swapping $l$ and $l'$ and  the properties of the delta function, we can further simplify the expression leading to 
\begin{equation}
    \begin{aligned}
    S_{\partial_i\phi \partial_j \phi}(\omega) &= \frac{\phi_0^2\omega}{4 (2\pi)^2} \int f(\vec{k}) k \sum_{ll'} \cos(\delta_l(k)-\delta_{l'}(k)) \\
    &\qquad\qquad\qquad\qquad \times\partial_i(\phi_l(k,r)P_l(\hat{k} \hat{r})) \partial_j(\phi_{l'}(k,r)P_{l'}(\hat{k} \hat{r})) d\Omega_{k}
    \end{aligned}
\end{equation}
where we have relabeled $\omega_0$ to $\omega$ for convenience and $k=\sqrt{\omega^2-m^2}$. Note that we have used the delta function to reduce the integral to 2-dimensional one over the solid angle $d\Omega_{k}$.

With a completely analogous derivation we can also obtain the power spectrum of the field value,
\begin{equation}
\label{eq:fieldpow}
    \begin{aligned}
    S_{\phi  \phi}(\omega) &= \frac{\phi_0^2\omega}{4 (2\pi)^2} \int f(\vec{k}) k \sum_{ll'} \cos(\delta_l(k)-\delta_{l'}(k)) \\
    &\qquad\qquad\qquad\qquad \times \phi_l(k,r)P_l(\hat{k} \hat{r}) \phi_{l'}(k,r)P_{l'}(\hat{k} \hat{r}) d\Omega_{k}~.
    \end{aligned}
\end{equation}

To evaluate these expressions we make use of the following simple velocity distribution in the Milky Way halo (cf., e.g.~\cite{disp}),
\begin{equation}
\label{eq:velocityspectrum}
f(\vec{k})=\frac{1}{(2\pi\sigma^2)^{3/2}}e^{-\frac{(\vec{k}-\vec{k}_0)^2}{2\sigma^2}}~,
\end{equation}
where $\sigma \sim 10^{-3}\, m$ is the typical momentum dispersion at the Earth's position and $k_0\sim 10^{-3}m$ is the shift in the distribution resulting from the movement of Earth with respect to the dark matter halo.

For calculational simplicity we choose a special point on Earth's surface for our measurements. We take Earth's movement in the z-direction of our coordinate system, with wave vector $\vec{k}_0 = (0,0,k_0)$, and put our measurement point in alignment, that is $\vec{r} = (0,0,R)$.  With this $P_l(\hat{k} \hat{r})=P_l(\cos(\theta))$ and the 2-dimensional integral simplifies to a single one.

In this case, using the expression $\phi_{\vec{k}}(\vec{r},t) = \Re \left( \phi_0\chi(\vec{k},\vec{r})e^{-i\omega t+i\alpha(\vec{k})} \right)$ for computational simplicity, the power spectra take the following form 
\begin{align}
    S_{\phi\phi}(\omega) &= \frac{\pi\phi_0^2\omega k}{2(2\pi)^\frac{7}{2}\sigma^3}\exp{\left(-\frac{k^2+k_0^2}{2\sigma^2}\right)}\int_0^\pi d\theta_k\sin{(\theta_k)}|\chi(\vec{k},\vec{r})|^2\exp{\left( \frac{kk_0}{\sigma^2}\cos{(\theta_k)}\right)}~, \\
    S_{\partial_i\phi\partial_j\phi}(\omega) &= \frac{\pi\phi_0^2\omega k}{2(2\pi)^\frac{7}{2}\sigma^3}\exp{\left(-\frac{k^2+k_0^2}{2\sigma^2}\right)}\int_0^\pi d\theta_k\sin{(\theta_k)}\partial_i\chi(\vec{k},\vec{r})\partial_j\chi^*(\vec{k},\vec{r})\exp{\left( \frac{kk_0}{\sigma^2}\cos{(\theta_k)}\right)}~.
\end{align}

To distinguish the effects of the interaction it will be useful to compare the "full" power spectrum, including the interaction, to the free power spectrum, for a plane wave $\chi =e^{i\vec{k}\cdot\vec{r}}$. For this plane wave we can determine the power spectrum analytically. For convenience
we give the solutions for the 2 most relevant power spectra $S_{\phi\phi}$ and $S_{\partial_r\phi\partial_r\phi}$,
\begin{align}
    S^{\rm{free}}_{\phi\phi}(\omega) \!&=\! \frac{\pi \phi_0^2}{(2\pi)^\frac{7}{2}}\frac{\omega}{\sigma k_0}\sinh{\left(\frac{kk_0}{\sigma^2} \right)}\exp{\left(-\frac{k^2 + k_0^2}{2\sigma^2} \right)}~, 
    \\
    S^{\rm{free}}_{\partial_r\phi\partial_r\phi} (\omega) \!&=\! \frac{\pi}{(2\pi)^\frac{7}{2}}\frac{\omega \sigma^3}{k_0^3}\left[( \left(\frac{kk_0}{\sigma^2}\right)^2+2)\sinh{\left(\frac{kk_0}{\sigma^2}\right)}-2\left(\frac{kk_0}{\sigma^2}\right)\cosh{\left(\frac{kk_0}{\sigma^2}\right)}\right]\exp{\left(-\frac{k^2 + k_0^2}{2\sigma^2} \right)}~.
\end{align}

\subsection{Results}
We are now ready to evaluate the effects  of a non-vanishing quadratic coupling on the power spectra as well as the effective changes in the signals expected for Earth bound experiments, both sensitive to gradients and the field value itself.

\begin{figure}[!t]    
\centering
    \includegraphics[width = 0.48 \linewidth]{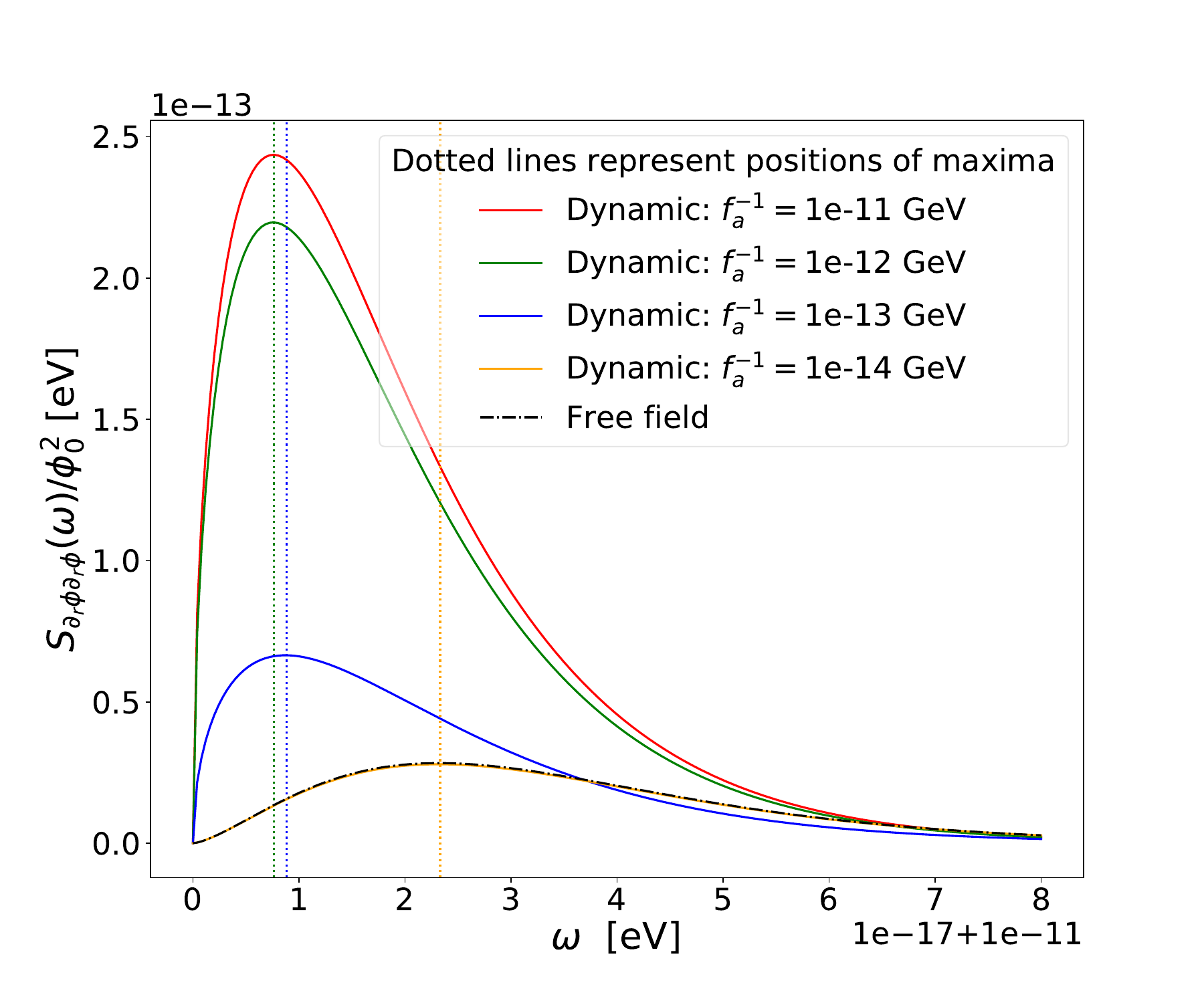}
    \includegraphics[width = 0.48 \linewidth]{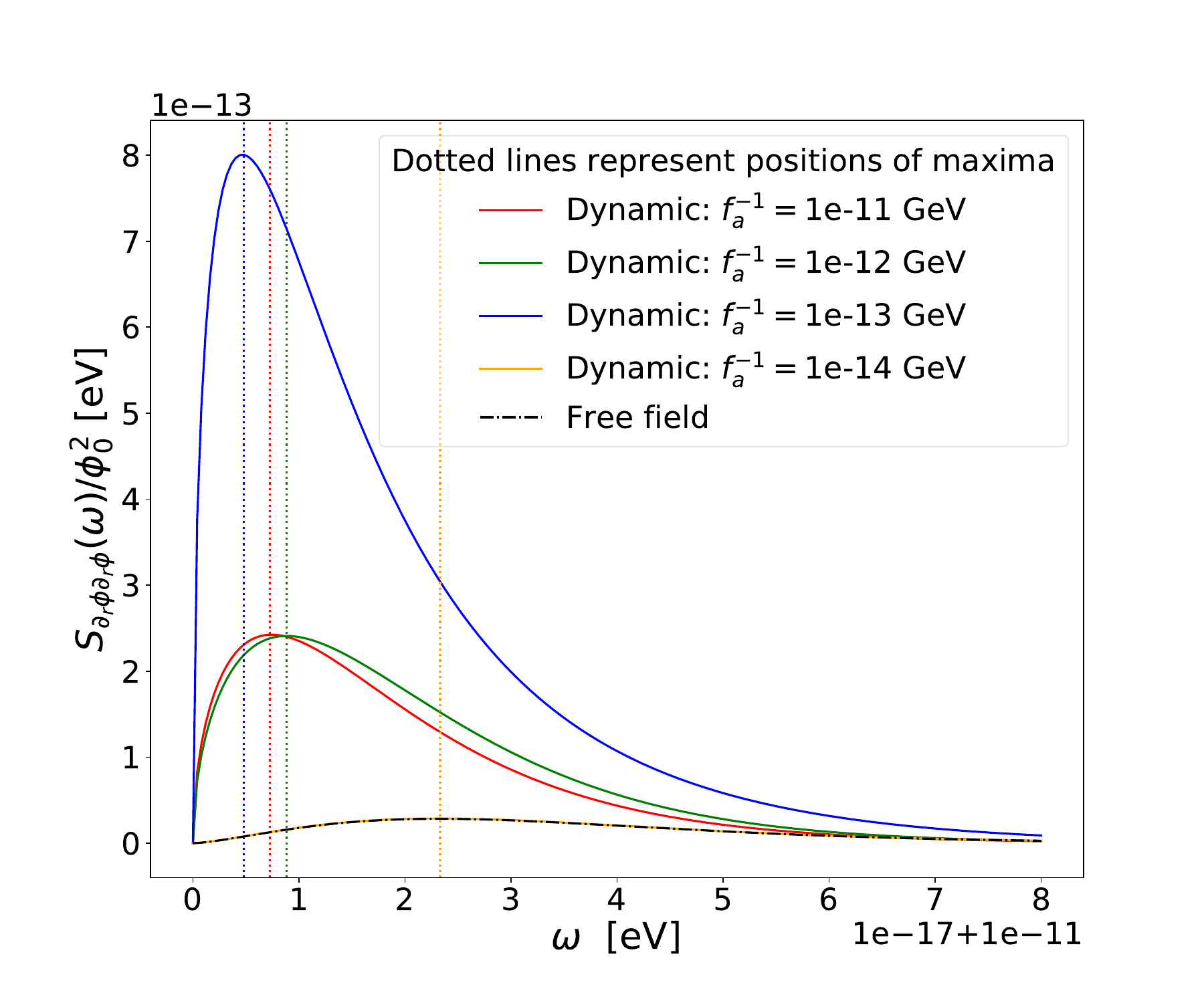}
    \caption{Changes in the power spectrum of the gradient for different couplings. The repulsive case is once again on the left and the attractive case on the right. We have chosen a mass of $m=10^{-11} \ \mathrm{eV}$ and correspondingly $\sigma = m\cdot 10^{-3}\,\mathrm{eV}$.  The vertical dashed lines give the positions of the maxima. The black dashed curve is the free field's radial power spectrum as reference.}
    \label{fig:gradient}
\end{figure}

\begin{figure}[!t]    
\centering
    \includegraphics[width = 0.48 \linewidth]{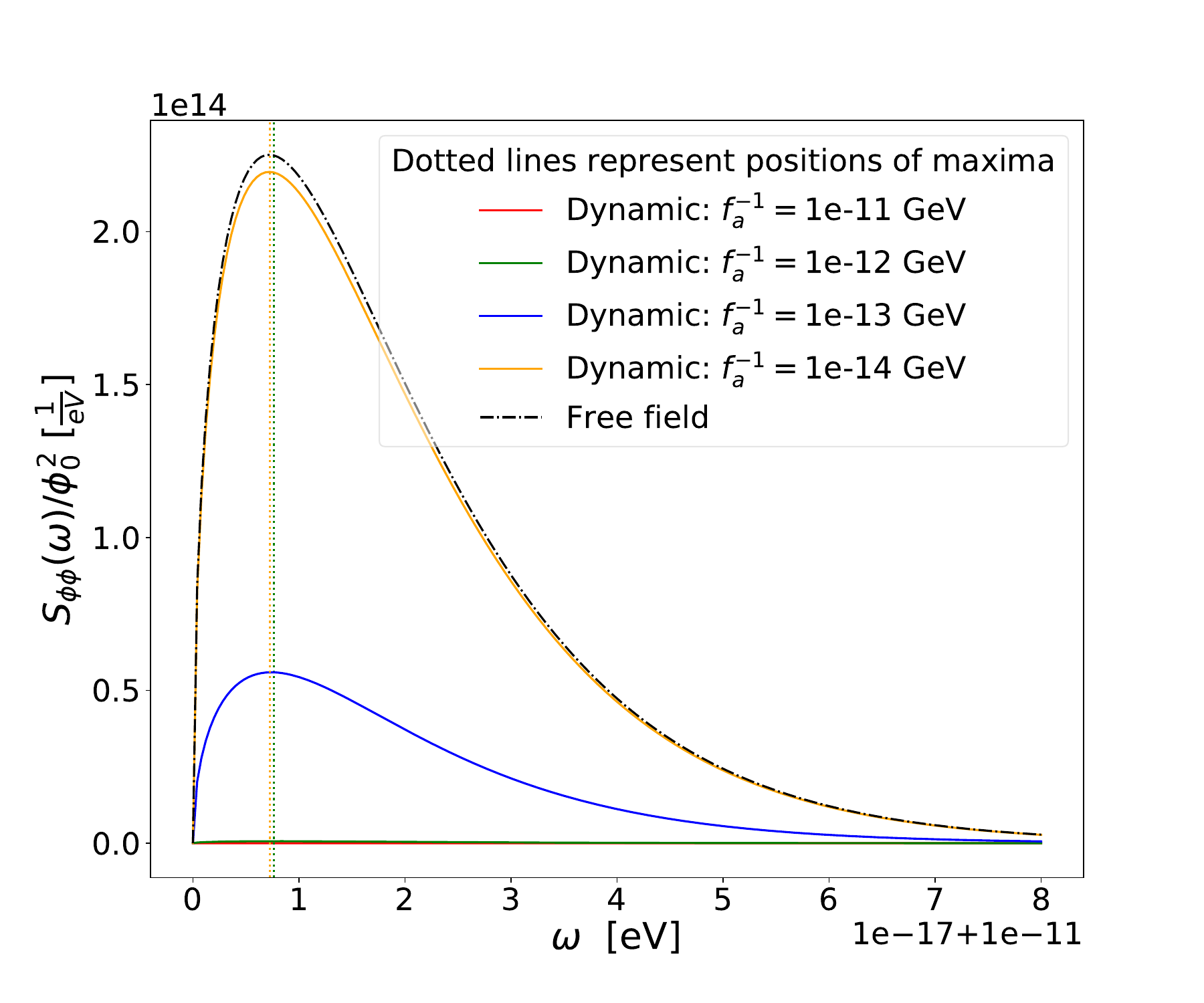}
        \includegraphics[width = 0.48 \linewidth]{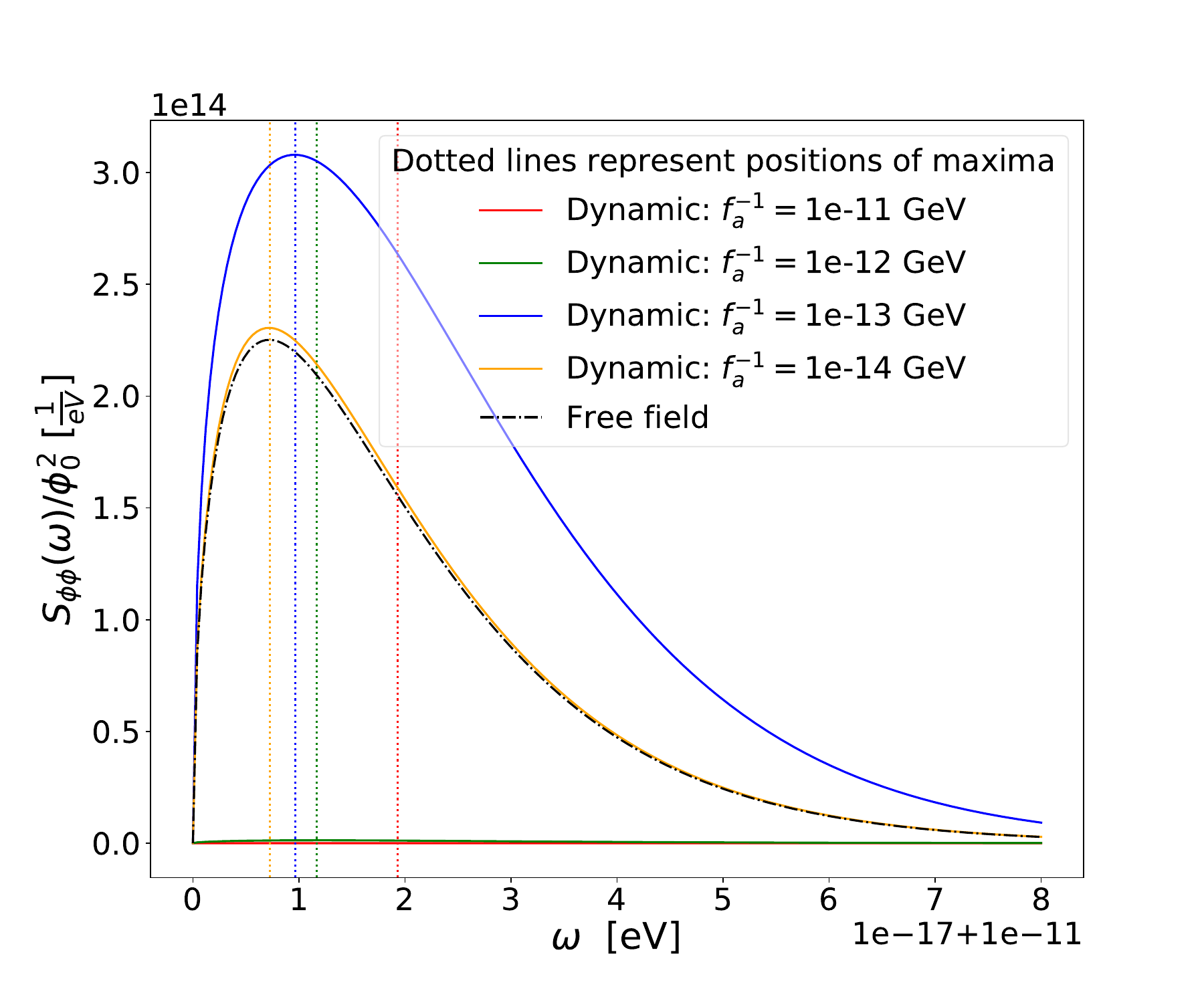}
    \caption{Changes in the power spectrum of the field for different couplings.  The repulsive case is once again on the left and the attractive case on the right.  We have chosen a mass of $m=10^{-11} \ \mathrm{eV}$ and correspondingly $\sigma = m\cdot 10^{-3}\mathrm{eV}$.  The vertical dashed lines give the positions of the maxima. The black dashed curve is the free field's field power spectrum as reference.}
    \label{fig:field}
\end{figure}

 In Fig.~\ref{fig:gradient} we show that the power spectrum of the gradient is modified for subsequently increasing couplings. We can see that the modifications first affects the low velocity modes, in agreement with the above argument about the balance between kinetic and potential energy. In Fig.~\ref{fig:field} we show the same for the field itself.

\begin{figure}[!t]    
\centering
    \includegraphics[width = 0.48 \linewidth]{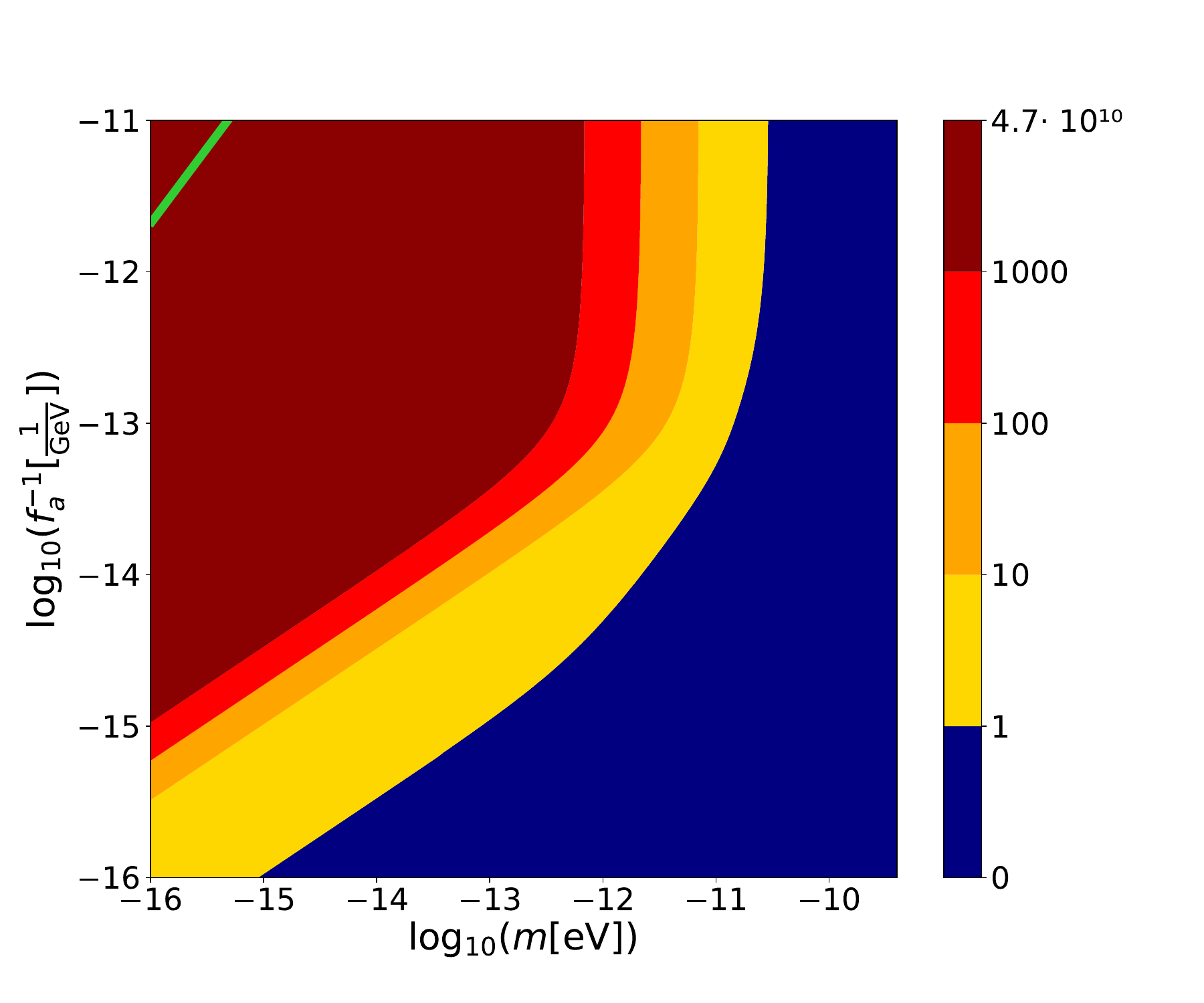}
    \includegraphics[width = 0.48 \linewidth]{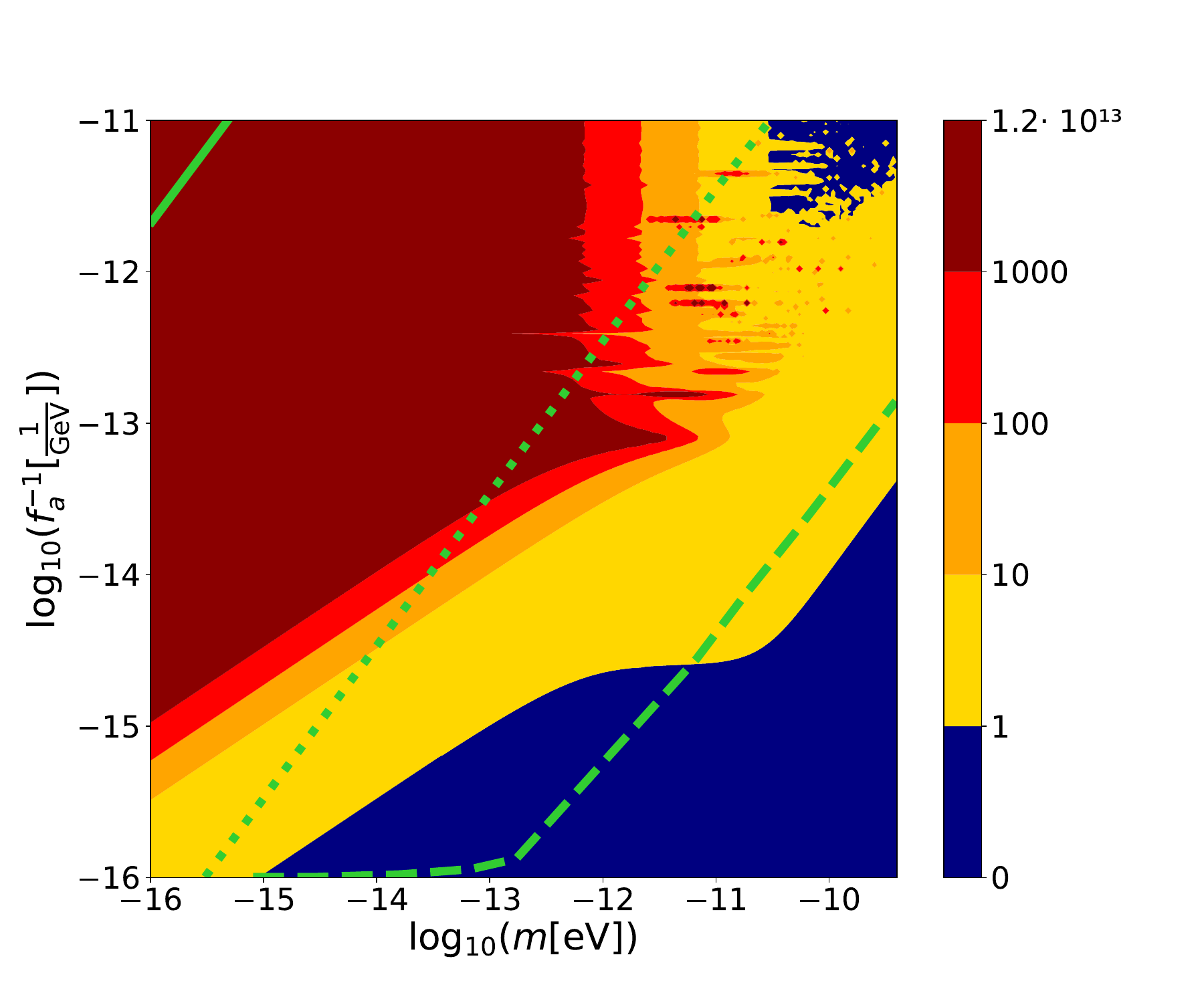}
    \caption{Effect of the quadratic coupling on the experimental sensitivities of experiments sensitive to the field gradient (left repulsive, right attractive). The bars to the right provide the color scheme for the ratio Eq.~\eqref{eq:powerratio}. All but the blue areas have enhanced signals. The regions above the green lines show existing constraints. The solid lines are based on experimental searches for dark matter~\cite{ Beam_EDM,HFH,I_2+Ca+,nEDM, ONIX,TriutiumDecay}. The dotted line corresponds to the BBN constraint~\cite{BBN} on dark matter coupled via gluons resulting in a change of the proton and neutron masses and we show it only in the attractive case. The dashed line is based on constraints on non-trivial static solutions~\cite{GW170817,NeutronStars_1,NeutronStars_2,WD,SolarCore,BinaryPulsar_Correction,BinaryPulsars} and again is shown only for the attractive case. The data for the limits was originally combined by~\cite{AxionLimits_OHare}. } 
    \label{fig:gradientpower}
\end{figure}

\begin{figure}[!t]    
\centering
    \includegraphics[width = 0.48 \linewidth]{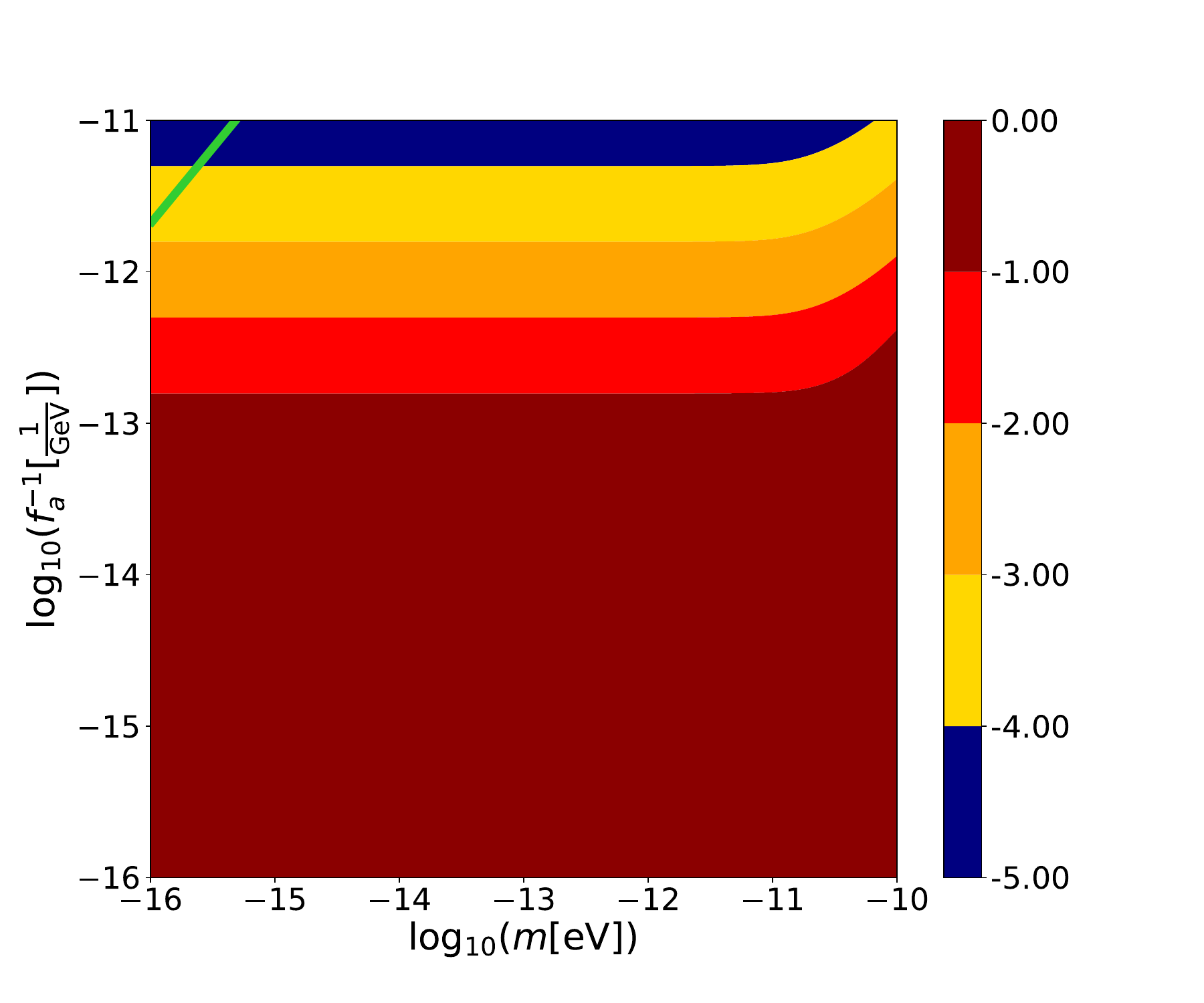}
    \includegraphics[width = 0.48 \linewidth]{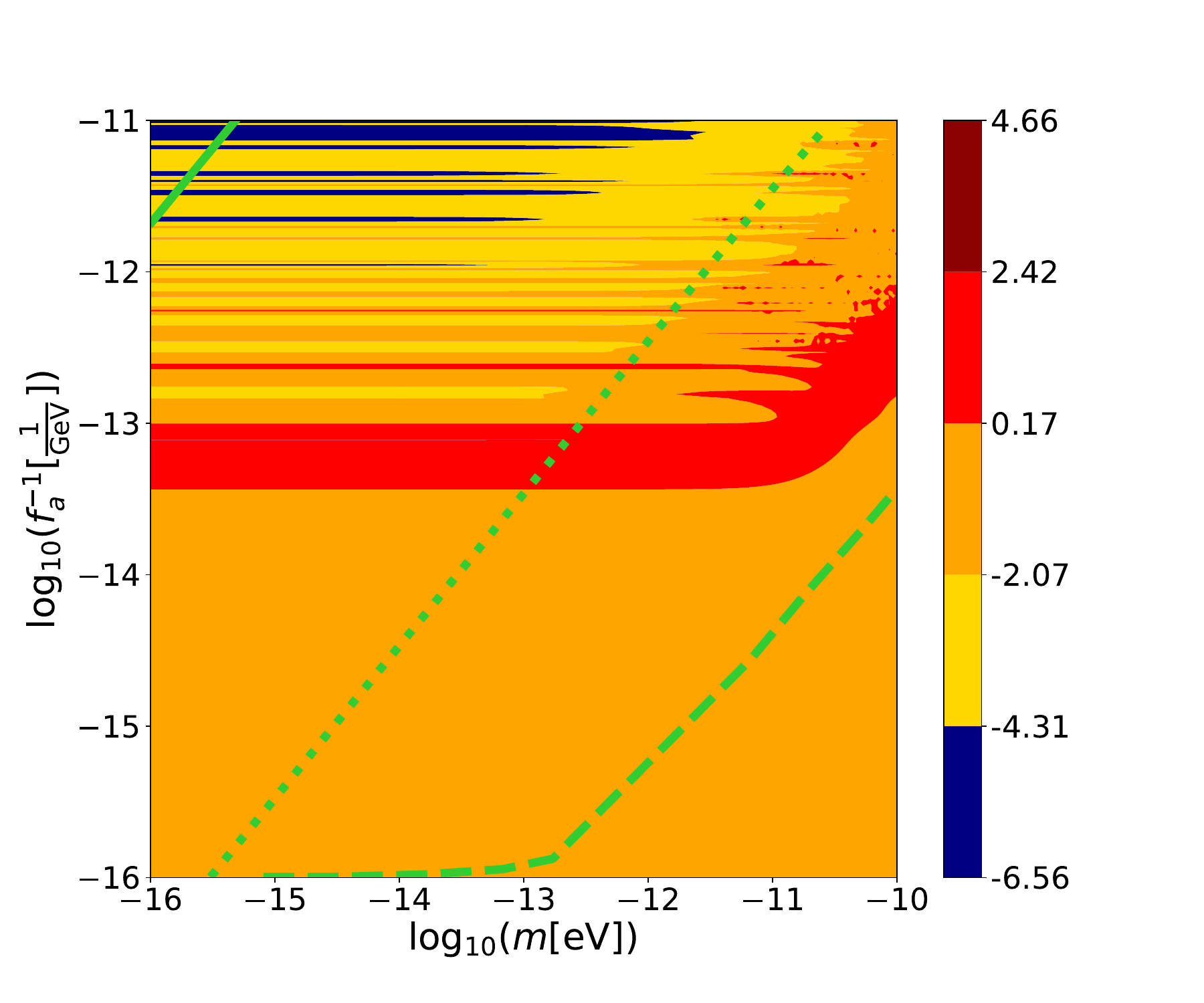}
    \caption{Effect of the quadratic coupling on the experimental sensitivities of experiments sensitive to the field itself (left repulsive, right attractive). The bars to the right provide the color scheme for the powers of 10 of the ratio Eq.~\eqref{eq:powerratio}. For the attractive case we note the same resolution issue as in figure \ref{fig:fieldcomparison}. The green lines for the constraints are as in Fig.~\ref{fig:gradientpower}, based on~\cite{BBN, Beam_EDM,HFH,I_2+Ca+,nEDM, ONIX,TriutiumDecay,GW170817,NeutronStars_1,NeutronStars_2,WD,SolarCore,BinaryPulsar_Correction,BinaryPulsars,AxionLimits_OHare}. }
    \label{fig:fieldpower}
\end{figure}

As we have just seen, different velocity modes are changed by different factors with respect to the case of vanishing quadratic coupling. Therefore, for a more quantitative determination of the effect of the quadratic coupling on the experimental sensitivity we have to take this effect into account. Using the power spectrum we can determine the overall change in  signal power compared to the case of vanishing coupling by appropriately integrating over frequency, 
\begin{equation}
\label{eq:powerratio}
    {\rm Ratio}=\frac{\int d\omega S_{X}(\omega)}{\int d\omega S^{\rm{free}}_{X}(\omega)}~.
\end{equation}
Here, $X$ denotes the relevant power spectrum in question, in particular we consider $X=\phi\phi$ and $X=\partial_{r}\phi\partial_{r}\phi$.
The results for the gradient and the field are shown in Figs.~\ref{fig:gradientpower} and \ref{fig:fieldpower}. While still involving approximations (in particular for the attractive case) and simple assumptions (e.g. we have evaluated the gradient power spectrum at a position directly in the ``wind'' of dark matter and considered the simple velocity spectrum of Eq.~\eqref{eq:velocityspectrum}) they should be considered our best estimates for the effect of a quadratic coupling on the sensitivity of experiments on the surface of the Earth. Therefore, we have also indicated the limits from astrophysical constraints as well as some exemplary experimental sensitivities from~\cite{BBN, Beam_EDM,HFH,I_2+Ca+,nEDM, ONIX,TriutiumDecay,GW170817,NeutronStars_1,NeutronStars_2,WD,SolarCore,BinaryPulsar_Correction,BinaryPulsars} based on the compilation of~\cite{AxionLimits_OHare}.

\subsection{Divergences in the attractive case}
As mentioned above and noted earlier in Refs.~\cite{Hees:2018fpg,Bauer:2024yow} the $v=0$ solution features divergences for certain attractive coupling values given by the condition $\sqrt{-\lambda\rho}=\frac{(2n+1)\pi}{2R}$ with $n$ integer. For attractive values of the coupling, Earth acts as an attractive potential well and the divergences appear whenever a new bound state becomes possible (see also the discussion in~\cite{lecture}). 
These divergences are absent for $v\neq0$, however they re-appear in the $k\to 0$ limit as a (cf.~Ref.~\cite{lecture} and Appendix~\ref{app:limit2})
\begin{equation}
\label{eq:divergencek}
    \phi\sim \phi_{0}\frac{1}{kR}
\end{equation}
behavior (on Earth's surface). 

As one can see from Eq.~\eqref{eq:fieldpow} (see also Appendix~\ref{app:limit2}) the power spectrum integrated over frequencies is, however, finite. 

That said, one may still wonder if there are potentially large and dangerous field values of the order of $\sim f_{a}$ as discussed in~\cite{Bauer:2024yow} in the $v=0$ case. Considering Eq.~\eqref{eq:divergencek} this may be particularly worrisome for small masses and correspondingly small $k\sim mv$.
Indeed, as also shown in Appendix~\ref{app:limit2}, the integrated power spectrum grows as,
\begin{equation}
    \int d\omega S(\omega)\sim \frac{1}{m^2} \qquad{\rm for }\,\, m\,\,{\rm small}.
\end{equation}
This again raises the issue of potentially large field values.

Therefore, let us estimate when this may become problematic. To do so we consider for which values of $k$ the field values may become problematically large.
For this we consider a single wave of momentum $k$ carrying the full dark matter density. We therefore have,
\begin{equation}
\phi_{0}\sim\frac{\sqrt{2\rho}}{m}~.
\end{equation}
Taking the coupling to be on resonance and taking the behavior shown in Eq.~\eqref{eq:divergencek} we find that this becomes problematic when,
\begin{equation}
    \phi\sim\phi_{0}\frac{1}{kR}\sim f_{a}~.
\end{equation}
This translates into (using $\rho=300\,{\rm MeV}/{\rm cm}^3$),
\begin{equation}
\label{eq:boundaryk}
k\sim \frac{\sqrt{2\rho}}{m}\frac{1}{Rf_{a}}\sim 7\times 10^{-20}\,{\rm eV}\left(\frac{10^{-16}\,{\rm eV}}{m}\right)\left(\frac{10^{10}\,{\rm GeV}}{f_{a}}\right).
\end{equation}
This small value indicates that these problems only occur for very small masses $\lesssim 10^{-16}\,{\rm eV}$.

Moreover, the problem is actually a time-dependent one (see also~\cite{Burrage:2024mxn}) and so far we have only considered the stationary solutions. One may wonder, at what time-scales we can actually distinguish between a mode with momentum just on the boundary Eq.~\eqref{eq:boundaryk} and a problematic one with $k=0$. Estimating the time it takes to accrue a phase difference of order $1$ between those two modes we find,
\begin{equation}
    t\sim \frac{2m}{k^2}\sim \frac{m^3}{\rho}(Rf_{a})^2
    \sim 1\,{\rm year}\left(\frac{m}{10^{-16}\,{\rm eV}}\right)^3\left(\frac{f_{a}}{10^{10}\,{\rm GeV}}\right)^2.
\end{equation}
On these time-scales the velocity of Earth with respect to dark matter changes by $\Delta v\sim 10^{-4}$ with respect to Earth due to the rotation of Earth around the sun.
This corresponds to a change in $k$,
\begin{equation}
    (\Delta k)_{\rm rotation}\sim 10^{-20}\,{\rm eV}\left(\frac{10^{-16}\,{\rm eV}}{m}\right) \qquad {\rm in\,\, one \,\, year}.
\end{equation}

We therefore think that for masses $m\gtrsim 10^{-16}\,{\rm eV}$ it is unlikely that field values $\sim f_{a}$ are reached. 
Nevertheless, a more careful study of the time-dependence may be interesting and shed light on values of $m\lesssim 10^{-16}\,{\rm eV}$.

\section{Conclusions}\label{sec:conclusions}
In this note we have studied the effect of a quadratic coupling to ordinary matter on the experimental sensitivities for very light bosonic dark matter. 
While at first sight such a quadratic interaction seems like an additional assumption we note that it is automatically present in many models, a notable example~\cite{Hook:2017psm,Bauer:2024yow} being gluon-coupled axion-like particles that are searched for in a variety of experiments.

Our main focus was on experiments sensitive to field gradients, e.g.~\cite{JacksonKimball:2017elr,QUAX:2020adt,Adelberger:2009zz,Graham:2015ifn}. We find that the field profiles induced by quadratic interactions with the matter in Earth can lead to a significant enhancement in the expected signal strength, particularly at low masses (cf. Fig.~\ref{fig:gradientpower}). Therefore, experiments may be more sensitive than in the naive case without quadratic interactions.

Using solutions of the field equations that allow for a non-vanishing velocity of the field with respect to Earth we can also improve upon the estimate of~\cite{Banerjee:2022sqg} for the change in sensitivity for experiments that are based on a coupling to the field itself. For higher masses the effect of quadratic interactions is significantly reduced compared to an approximation that neglects the incoming velocity.
This is because the kinetic energy of the particles is large compared to the potential caused by the quadratic interaction. The results for a full velocity distribution are shown in Fig.~\ref{fig:fieldpower}.

For attractive interactions and vanishing velocities the solutions for vanishing velocity diverge for certain values of the coupling (see also~\cite{Hees:2018fpg,Bauer:2024yow}). We think that these problems are related to resonances caused by the appearance of bound states. For non-vanishing velocities these divergences are absent. This also remains true for the frequency integrated power spectrum. Nevertheless, for very small masses $m\lesssim 10^{-16}\,{\rm eV}$ problematically large field values may still appear. This issue suggests further investigation in a time-dependent context.

All in all quadratic interactions with ordinary matter, of a size that one would expect alongside the typical linear interactions used for experiments, can lead to sizable changes in the field values and gradients that determine the sensitivity of dark matter experiments performed here on Earth. Therefore, their effect should be considered when delineating the sensitivity.

\section*{Acknowledgements}
We would like to thank Clare Burrage and Benjamin Elder for interesting discussions and collaboration on related topics. We also thank the authors of ~\cite{Banerjee:2025dlo} for coordinating the release of our works which aim at similar physics. Parts of this work were performed in the context of and are based on the Master thesis of YGdC and the Bachelor thesis of BH. JJ would like to thank the EU for support via ITN HIDDEN (No 860881).
\appendix
\section{Continuity constants of stationary solutions}\label{app:continuity}
In this appendix we provide the continuity constants of the stationary solutions to the  field equations when the density is given by Eq. \eqref{2layerDensity} (with the second layer removed for the one layer model). The solutions are provided in equations \eqref{In1stLayer}, \eqref{In2ndLayer} and \eqref{OutsideBothLayers}.
\newline
We can determine the constants from the continuity conditions of the field and its radial derivative at the layer's boundaries. In the two layer model there are four constants A, B, C and D. They are given by
\begin{align}
\label{eq:twolayerrep}
    A &= \frac{\sinh(\gamma_2(R_1-R_2))[R_2\gamma_2\sinh(\gamma_1R_1)-\frac{\gamma_1}{\gamma_2}\cosh(\gamma_1R_1)]}{\gamma_2\sinh(\gamma_1R_1)\sinh(\gamma_2(R_1-R_2))-\gamma_1\cosh(\gamma_1R_1)\cosh(\gamma_2(R_1-R_2))}\\ \nonumber
    &+\frac{\cosh(\gamma_2(R_1-R_2))[\sinh(\gamma_1R_1)-R_2\gamma_1\cosh(\gamma_1R_1)]}{\gamma_2\sinh(\gamma_1R_1)\sinh(\gamma_2(R_1-R_2))-\gamma_1\cosh(\gamma_1R_1)\cosh(\gamma_2(R_1-R_2))} \\ \nonumber
    \\
    B &= \frac{1}{\gamma_1\cosh(\gamma_1R_1)\cosh(\gamma_2(R_1-R_2))-\gamma_2\sinh(\gamma_1R_1)\sinh(\gamma_2(R_1-R_2))}\\ \nonumber
    \\
    C &= \frac{\frac{1}{2}[\sinh(\gamma_1R_1)+\frac{\gamma_1}{\gamma_2}\cosh(\gamma_1R_1)]\exp(-\gamma_2R_1)}{\gamma_1\cosh(\gamma_1R_1)\cosh(\gamma_2(R_1-R_2))-\gamma_2\sinh(\gamma_1R_1)\sinh(\gamma_2(R_1-R_2))}\\ \nonumber
    \\ 
    D &= \frac{\frac{1}{2}[\sinh(\gamma_1R_1)-\frac{\gamma_1}{\gamma_2}\cosh(\gamma_1R_1)]\exp(\gamma_2R_1)}{\gamma_1\cosh(\gamma_1R_1)\cosh(\gamma_2(R_1-R_2))-\gamma_2\sinh(\gamma_1R_1)\sinh(\gamma_2(R_1-R_2))}
\end{align}
for $\lambda>0$ and
\begin{align}
\label{eq:twolayeratt}
    A &= \frac{\sin(\gamma_2(R_1-R_2))[R_2\gamma_1\cos(\gamma_1R_1)-\sin(\gamma_1R_1)]}{\gamma_2\sin(\gamma_2(R_1-R_2))\sin(\gamma_1R_1)+\gamma_1\cos(\gamma_1R_1)\cos(\gamma_2(R_1-R_2))}\\ \nonumber
    &+\frac{\cos(\gamma_2(R_1-R_2))[R_2\gamma_2\sin(\gamma_1R_1)+\frac{\gamma_1}{\gamma_2}\cos(\gamma_1R_1)]}{\gamma_2\sin(\gamma_2(R_1-R_2))\sin(\gamma_1R_1)+\gamma_1\cos(\gamma_1R_1)\cos(\gamma_2(R_1-R_2))}\\ \nonumber
    \\
    B &=\frac{1}{\gamma_2\sin(\gamma_2(R_1-R_2))\sin(\gamma_1R_1)+\gamma_1\cos(\gamma_1R_1)\cos(\gamma_2(R_1-R_2))} \\ \nonumber
    \\
    C &=\frac{\sin(\gamma_1R_1)\sin(\gamma_2R_1)+\frac{\gamma_1}{\gamma_2}\cos(\gamma_1R_1)\cos(\gamma_2R_1)}{\gamma_2\sin(\gamma_2(R_1-R_2))\sin(\gamma_1R_1)+\gamma_1\cos(\gamma_1R_1)\cos(\gamma_2(R_1-R_2))} \\ \nonumber
    \\
    \label{eq:dequation}
    D &=\frac{\sin(\gamma_1R_1)\cos(\gamma_2R_1)-\frac{\gamma_1}{\gamma_2}\cos(\gamma_1R_1)\sin(\gamma_2R_1)}{\gamma_2\sin(\gamma_2(R_1-R_2))\sin(\gamma_1R_1)+\gamma_1\cos(\gamma_1R_1)\cos(\gamma_2(R_1-R_2))} 
\end{align}
for $\lambda<0$. The one layer model has only two constants, A and B, given by
\begin{align}
\label{eq:onelayerrep}
    B & = \frac{1}{\cosh{\left( R_1\gamma_1 \right)}\gamma_1}\\ \nonumber
    A & = R_1^3\gamma_1^2\frac{R_1\gamma_1-\tanh{(R_1\gamma_1)}}{(R_1\gamma_1)^3} \nonumber
\end{align}
for $\lambda>0$ and
\begin{align}
\label{eq:onelayeratt}
    B & = \frac{1}{\cos{\left( R_1\gamma_1 \right)}\gamma_1}\\ \nonumber
    A & = R_1^3\gamma_1^2\frac{R_1\gamma_1-\tan{(R_1\gamma_1)}}{(R_1\gamma_1)^3} \nonumber
\end{align}
for $\lambda<0$.

\section{Limit of vanishing velocity}\label{app:limit}
 We expect that the solutions we have found for non-vanishing velocity, Eqs.~\eqref{eq11} and \eqref{eq22}, in the limit $k \rightarrow 0$, coincide with the non-moving one layer solutions given by taking the limit of vanishing atmosphere, i.e. $R_{1}=R_{2}=R$ of Eq.~\eqref{In1stLayer},  \eqref{In2ndLayer} and \eqref{OutsideBothLayers}. The latter also agree with the single layer solution explicitly given in Ref.~\cite{Hees:2018fpg}.

Taking for example the solution given by Eq.~\eqref{eq11}, we only need to take into account the $l=0$ term, since we are taking the limit to a homogeneous boundary condition. The limit for $r<R$ then reads (for simplicity we choose the phase of the incoming dark matter wave to be $\delta=0$).
\begin{eqnarray}
\nonumber
\lim_{k \to 0}\phi(\vec{r},t)&&
\\\nonumber
&&\!\!\!\!\!\!\!\!\!\!\!\!\!\!\! \!\!\!=\lim_{k \to 0} \Re \left( \phi_{0} e^{-imt} \left(\frac{j_0(kR)}{j_0(k'R)} +\frac{k R j_0(k'R)j_0'(kR)-j_0'(k'R)j_0(kR)}{h_0(kR) k'R j_0(k'R) j_0'(k'R)-k R j_0(k'R) h_0'(k R)})j_0(k'r) \right) \right) \nonumber \\
&&\!\!\!\!\!\!\!\!\!\!\!\!\!\!\! \!\!\!=\lim_{k \to 0} \phi_{0} \cos(mt) \left(\frac{1}{j_0(k'R)}+\frac{-j_0'(k'R)+\mathcal{O}(kR)}{k'R j_0(k'R) j_0'(k'R)+ j_0(k'R)^2+\mathcal{O}(kR)})j_0(k'r) \right) \nonumber \\
&&\!\!\!\!\!\!\!\!\!\!\!\!\!\!\! \!\!\!=\phi_{0} \cos(mt) \left(\frac{1}{j_0(\gamma R)+\gamma R j_0'( \gamma R)}\frac{\sin(\gamma r)}{\gamma r} \right) \nonumber \\
&&\!\!\!\!\!\!\!\!\!\!\!\!\!\!\! \!\!\!=\phi_{0} \cos(mt) \left(\frac{1}{\cos(\gamma R)}\frac{\sin(\gamma r)}{\gamma r} \right)~,
\end{eqnarray}
where $\gamma=\sqrt{|\lambda| \rho}$ and the last expression matches with Eq.~\eqref{In1stLayer} in the limit of vanishing atmosphere. For $r>R$, the limit reads,
\begin{eqnarray}
\nonumber
\lim_{k \to 0}\phi(\vec{r},t)\!\!\!&=&\!\!\! \lim_{k \to 0} \Re \left( \phi_{0} e^{-imt} (j_0(kr)+ \frac{k j_{0}'(k R)j_{0}(k' R)- k'j_{0}'(k' R)j_{0}(k R)}{k' h_{0}(k R)j_{0}'(k' R)-k j_{0}(k 'R)h_{0}'(k R)}h_0(kr)) \right)
\\\nonumber
\!\!\!&=&\!\!\!\lim_{k \to 0}  \phi_{0} \cos(mt)\left(1-\frac{ k'j_{0}'(k' R)}{\frac{1}{k R}k'j_{0}'(k' R)+ \frac{1}{k R^2}j_{0}(k 'R) }\frac{1}{k r}\right)
\\ \nonumber
\!\!\!&=&\!\!\! \phi_{0} \cos(mt)\left( 1-\frac{\gamma R j_0'(\gamma R)}{j_0(\gamma R)+\gamma Rj_0'(\gamma R)} \frac{R}{r} \right) 
\\
\!\!\!&=&\!\!\! \phi_{0} \cos(mt)\left( 1-R^3\gamma^2\frac{R\gamma-\tan{(R\gamma)}}{(R\gamma)^3}\frac{1}{r} \right) ~.
\end{eqnarray}

This is exactly what was obtained in Eq.~\eqref{OutsideBothLayers} for $r>R$. The same kind of limit can be done for the repulsive case and the result is analogous.
\section{Proof that the attractive dynamic solutions contain no divergences}\label{Proof}
In the attractive case we only need to concern ourselves with the case $\omega^2-m^2-\lambda\rho>0$ which, as a reminder, is given by
\begin{equation} 
    \phi(\vec{r},t)= 
    \begin{cases}
   \Re \left( \phi_{0} e^{-i \omega t} \sum_{l=0}^{\infty} A_l j_l(k' r)P_l(\cos(\theta))\right)& \text{if } r<R\\
     \Re \left( \phi_{0} e^{-i \omega t}  \sum_{l=0}^{\infty}(i)^l (2l+1)(j_l(kr)+ia_l h_l^1(kr))P_l(\cos(\theta)) \right)& \text{if }  r>R~.
    \end{cases}
\end{equation}
These can only diverge when the continuity constant 
\begin{equation}
i a_{l}=\frac{k j_{l}'(k R)j_{l}(k' R)- k'j_{l}'(k' R)j_{l}(k R)}{k' h_{l}(k R)j_{l}'(k' R)-k j_{l}(k 'R)h_{l}'(k R)}
\end{equation}
diverges\footnote{We have explicitly checked that divergences of $A_l$ only arise when $a_l$ diverges.}. Its poles are given by the zeros of
\begin{equation}\label{DangerousDenominator}
    k' h_{l}(k R)j_{l}'(k' R)-k j_{l}(k 'R)h_{l}'(k R)~.
\end{equation}
We can already exclude those zeros, where both $j_{l}(k' R)$ and $j_{l}'(k' R)$ are zero at the same time, as they would cancel with the numerator. We will show, that for $k>0 \implies k'>0$ and $0\leq l$, which are always satisfied if there is a non vanishing initial velocity, the above expression has no zeros, and the solutions therefore don't contain poles. To do this we will need 3 properties of the spherical Bessel functions of first and second kind
\begin{enumerate}
    \item $f'_{n+1}(x) = f_n(x)-\frac{n+2}{x}f_{n+1}(x)$ \label{1stProperty} \\
    \item $f_{n+1}(x) = \frac{n}{x}f_n(x)-f'_n(x)$\label{2ndProperty} 
    \item For $0\leq n$ the zeros of the Bessel functions of first and second kind of the same order are interlaced. As a result they have no common zeros. The same is true for spherical Bessel functions of different orders \cite{NIST:DLMF}. \label{3rdProperty}
\end{enumerate}
Here $f_n$ may refer to either the spherical Bessel function of first ($j_n$) or second kind  ($y_n$).
\begin{proof}
Setting \eqref{DangerousDenominator} to zero, and expanding the Hankel functions, we find two conditions because the denominator is complex but the functions $j_n$ and $y_n$ and their derivatives are real for real inputs, so the imaginary and real parts must vanish independently. Combining both conditions we arrive at
\begin{equation}\label{ToBeDisproven}
    \frac{j_n'(x)}{j_n(x)} = \frac{y_n'(x)}{y_n(x)}
\end{equation}
where we defined $x \equiv kR$. Now showing that this relation does not hold implies the result we want, since $(A \implies B)\Leftrightarrow(\neg B \implies \neg A)$. We will do this by induction, for $n = 0$, we have $j_n(x) = \frac{\sin(x)}{x}$ and $y_n = -\frac{\cos(x)}{x}$. Then \eqref{ToBeDisproven} gives us
\begin{equation}
    \cot(x) = -\tan(x)~.
\end{equation}
Since $\cot(x)$ and $\tan(x)$ always have the same sign and share no real zeros, this is impossible and we move on to the induction hypothesis. Assuming \eqref{ToBeDisproven} to be false for some integer $n$ we show that the equation is wrong for $n+1$. Starting from
\begin{equation}
    \frac{j_{n+1}'(x)}{j_{n+1}(x)} = \frac{y_{n+1}'(x)}{y_{n+1}(x)}
\end{equation}
we can first use \ref{1stProperty}, which gives us
\begin{equation}
    -(n+2) + \frac{1}{x}\frac{j_n(x)}{j_{n+1}(x)} = -(n+2) + \frac{1}{x}\frac{y_n(x)}{y_{n+1}(x)}
\end{equation}
for $x>0$ we can cancel all but the quotients
\begin{equation}
    \frac{j_n(x)}{j_{n+1}(x)} = \frac{y_n(x)}{y_{n+1}(x)}~.
\end{equation}
Now, we use \ref{3rdProperty}, since we know that $y_n$ and $j_n$ cannot be zero at the same time, we can invert the above expressions and then use \ref{2ndProperty} on $j_{n+1}$ and $y_{n+1}$ (we also do not need to worry about $j_{n+1}$ and $j_n$ being 0 at the same time, complicating things, as this is also impossible because of~\ref{3rdProperty}), this yields
\begin{equation}
    -\frac{n}{x}+\frac{j'_n(x)}{j_n(x)} = -\frac{n}{x}+\frac{y'_n(x)}{y_n(x)}~.
\end{equation}
Of course, now, we can simply use our induction hypothesis, that this expression is incorrect, implying that the expression $\frac{j_{n+1}'(x)}{j_{n+1}(x)} = \frac{y_{n+1}'(x)}{y_{n+1}(x)}$ is false, completing our proof by induction. 
\end{proof}

\section{Some comments on divergences}\label{app:limit2}

 In Appendix \ref{Proof} we showed that the dynamical solutions do not present divergences as long as $k \neq 0$. However one could in principle expect divergences in the integrated power spectrum where an integral over all frequencies is carried out, as this usually includes a region with vanishing relative velocity. In order to asses whether this happens or not, it is useful to study the behavior of the the dynamical solutions in the $k \to 0$ limit. 
 
 Moreover, in Appendix \ref{app:limit} we already showed that dynamical solutions reduce to the non-moving ones in the $k \to 0$ limit. It is obvious that if the non-moving solutions are finite the power spectrum will also be. However, as we already mentioned in the main text, the attractive non-moving solutions feature divergences for values of the coupling that fulfill $\sqrt{-\lambda\rho}=\frac{(2n+1)\pi}{2R}$, and we should study these cases more carefully. Taking for simplicity the first critical coupling $\lambda=-\frac{\pi^2}{4R^2\rho}$, the dynamical solution for $r>R$ (Eq. \eqref{eq11}) reads

\begin{multline}
\phi(\vec{r},t) = \Re \Bigg( \phi_o e^{-i\omega t} \frac{e^{-i k R}}{k r} \Bigg[ 
kR \sin \left(\sqrt{(kR)^2+\frac{\pi^2}{4}}\right) \cos (kR - kr) \\
- \sqrt{(kR)^2+\frac{\pi^2}{4}} \cos \left(\sqrt{(kR)^2+\frac{\pi^2}{4}}\right) \sin (kR - kr) \Bigg] \Bigg/ \Bigg[ 
\sqrt{(kR)^2+\frac{\pi^2}{4}} \cos \left(\sqrt{(kR)^2+\frac{\pi^2}{4}}\right) \\
- i kR \sin \left(\sqrt{(kR)^2+\frac{\pi^2}{4}}\right) \Bigg] \Bigg)~,
\end{multline}
where we have only taken into account the term with $l=0$ since it is the dominant one in the $k\to0$ limit. Expanding this expression around $k=0$ we arrive to

\begin{equation}
    \phi(\vec{r},t) = \Re \left(  \phi_o e^{-i\omega t}(\frac{i}{kr}+\frac{R}{2r}+\mathcal{O}(k) )\right)\approx \phi_0\cos(\omega t-\pi/2)\frac{1}{kr}~,
    \label{series}
\end{equation}
which is a good approximation if $k\ll\frac{1}{R}$. Note also 
 that there is a phase shift of $\pi/2$ which is expected to appear for resonances (see also Ref.~\cite{lecture}). From this expansion we can infer that the dynamical solutions diverge as $\sim 1/k$ for the critical couplings.

With this information, we can already see that the power spectrum given by Eq.~\eqref{eq:fieldpow} is going to diverge when $\omega\to m$ ($v\to 0$) and $\sqrt{-\lambda \rho}=\frac{(2n+1)\pi}{2R}$. The divergence for $r=R$ goes as, 
\begin{eqnarray}
    S_{\phi  \phi}(\omega\to m) &\approx& \frac{\phi_0^2\omega}{4 (2\pi)^2} \int \frac{k}{(2\pi\sigma^2)^{3/2}}e^{-\frac{k_0^2}{2\sigma^2}} \frac{1}{kR}\frac{1}{kR} d\Omega_{k}~
\\\nonumber
&\approx& \frac{\phi_0^2 e^{-\frac{k_0^2}{2\sigma^2}}}{2(2\pi)^{5/2}\sigma^3R^2}\frac{\omega}{k}~
\\\nonumber
&\approx& \frac{\phi_0^2 e^{-\frac{k_0^2}{2\sigma^2}}}{2(2\pi)^{5/2}\sigma^3R^2}\frac{1}v{}\sim1/v~.
\end{eqnarray}
On the other hand, the integrated power spectrum does not diverge even for critical couplings. Taking the field version of Eq.~\eqref{powdelta}, the integrated power spectrum reads,
\begin{eqnarray}
 \int d\omega S_{\phi \phi}(\omega)&\approx& \int d\omega\frac{\phi_0^2}{4 (2\pi)^2} \int f(\vec{k}) \delta(\omega_0-\omega)\phi_0(k,R)\phi_{0}(k,R)d^3k
\\\nonumber
 &=& \frac{\phi_0^2}{4 (2\pi)^2} \int  \frac{1}{(2\pi\sigma^2)^{3/2}}e^{-\frac{(\vec{k}-\vec{k}_0)^2}{2\sigma^2}} \frac{1}{kR} \frac{1}{kR}d^3k~,
\end{eqnarray}
where we have assumed that $k_0\ll1/R$ and, in consequence, the field is well approximated by Eq.~\eqref{series} for the dominant modes. In this regime, we can also assume this behavior for higher $k$ modes since they will be very suppressed by the momentum distribution and the resulting integrated power spectrum should not change significantly. Choosing $\vec{k}_0$ in the z direction, the integral simplifies to,
\begin{eqnarray}
\int d\omega S_{\phi \phi}(\omega)&\approx&  \frac{\phi_0^2}{4 (2\pi)^{5/2}\sigma^3R^2} e^{-\frac{k_0^2}{2\sigma^2}}\int e^{-\frac{k^2}{2\sigma^2}}  dk \int e^{\frac{k k_0\cos(\theta)}{\sigma^2}}d(\cos(\theta))
\\\nonumber 
&=& \frac{\phi_0^2e^{-\frac{k_0^2}{2\sigma^2}}  \text{erfi}\left(\frac{k_0 }{\sqrt{2}\sigma}\right)}{4(2\pi)^{3/2}R^2} \frac{1}{\sigma k_0}\sim1/m^2~,
\end{eqnarray}
which shows that the integrated power spectrum is always finite. That said, we can also see that the power spectrum grows with decreasing mass, potentially opening the possibility for problematically large field values. We discuss this in the main text.

If $k_0\not\ll1/R$ we can also show that the result is finite. Since we  know that the possible source of divergence comes from the power spectrum at $\omega=m$ ($k=0$), in order to asses whether the integrated one diverges or not, we can restrict the integration domain from $k=0$ to a certain cut-off $k=\Lambda$ which fulfills $\Lambda\ll k_0$ and $\Lambda\ll 1/R$. This cut-off should also be understood to be proportional to the mass in order to fulfill $\Lambda\ll k_0$ for all masses. In this case the integrated power spectrum over this restricted domain reads
\begin{eqnarray}
 \int_m^{\sqrt{m^2+\Lambda^2}} d\omega S_{\phi \phi}(\omega)&\approx&\frac{\phi_0^2 \pi}{(2\pi)^{7/2}\sigma^3R^2} e^{-\frac{k_0^2}{2\sigma^2}}\int_0^\Lambda  dk
\\\nonumber 
&=&\frac{\phi_0^2 e^{-\frac{k_0^2}{2\sigma^2}}}{ 2(2\pi)^{5/2}R^2}\frac{ \Lambda}{\sigma^3}\sim1/m^2~,
\end{eqnarray}
which, again, is a finite result.

\bibliographystyle{utphys}
\bibliography{references}

\end{document}